\documentclass{aa}  

\usepackage{hyperref}
\usepackage{graphicx}
\usepackage{tabularx}
\usepackage{xcolor}
\usepackage{txfonts}
\usepackage{xspace}
\usepackage{orcidlink}
\usepackage{caption}

\def\HI{\ion{H}{I}}

\newcommand{\CO}{\element[][12]CO}
\newcommand{\couno}{\element[][12]CO (1--0)\,}
\newcommand{\codue}{\element[][12]CO (2--1)\,}
\newcommand{\cotre}{{\element[][12]CO (3--2)}\,}

\usepackage{printlen}
\usepackage{layouts}

\newcommand{\kms}{$\,$km~s$^{-1}$}

\newcommand{\Jykms}{Jy\,km s$^{-1}$}
\newcommand{\mJyb}{mJy\,beam$^{-1}$}

\newcommand{\whz}{W~Hz$^{-1}$}
\newcommand{\msun}{{${\rm M}_\odot$}}
\newcommand{\msunyr}{{${\rm M}_\odot$ yr$^{-1}$}}

\newcommand{\cmsq}{cm$^{-2}$}

\newcommand{\eg}{\mbox{e.g.}}
\newcommand{\ie}{\mbox{i.e.}}

\newcommand{\forn}{\mbox Fornax~A}
\newcommand{\ntr}{\mbox NGC~3100}

\newcommand{\nzr}{\mbox NGC~3095}

\newcommand{\meer}{{MeerKAT}}

\newcommand{\cara}{{\tt CARACal}}

\newcommand{\vsys}{{$v_{\rm sys}$}\,}

\newcommand{\nhi}{{$N_{\rm H {\hskip 0.02cm \tt I}}$}}
\newcommand{\mhi}{{$M_{\rm H {\hskip 0.02cm \tt I}}$}}

\newcommand{\magn}{MAGNHIFFIC}

\begin{document} 

   \title{A \magn\ cold-gas filament fuelling the nuclear activity of \mbox{NGC 3100}}
   \titlerunning{\meer\ HI\ observations of \ntr}

        \author{F.~M.~Maccagni\inst{1,2}\corrauth{filippo.maccagni@inaf.it}
        \and 
        I.~Prandoni\inst{3,4}\email{isabella.prandoni@inaf.it}\and  
        F.~Fonzo\inst{5}\email{francesca.fonzo@inaf.it}\and  
        J.~Healy\inst{6,4}\email{julia.healy@manchester.ac.uk}\and   
        A.~Loni\inst{5}\email{alessandro.loni@inaf.it}\and  
        R.~Ragusa\inst{5}\email{rossella.ragusa@inaf.it}\and   
        I.~Ruffa\inst{7}\email{ilaria.ruffa@inaf.it}\and    
        W.~J.~G.~de~Blok\inst{8,9,4}\email{blok@astron.nl}\and    
        M.~Gaspari\inst{10}\email{massimo.gaspari@unimore.it}\and   
        P.~Serra\inst{1}\email{paolo.serra@inaf.it}\and   
        F.~Barbani\inst{10}\email{filippo.barbani@unimore.it}\and   
        E.~Iodice\inst{5}\email{enrichetta.iodice@inaf.it}\and
        D.~Kleiner\inst{8}\email{kleiner@astron.nl}\and   
        M.~Spavone\inst{5}\email{marilena.spavone@inaf.it}  }

   \institute{
        INAF -- Osservatorio Astronomico di Cagliari, via della Scienza 5, 09047, Selargius (CA), Italy
        \and
        Wits Centre for Astrophysics, School of Physics, University of the Witwatersrand, 1 Jan Smuts Avenue, 2000, Johannesburg, South Africa
        \and
        INAF -- Istituto di Radioastronomia, via P.\ Gobetti 101, 40129 Bologna, Italy
        \and
        Dept.\ of Astronomy, Univ.\ of Cape Town, Private Bag X3, Rondebosch 7701, South Africa
        \and
        INAF -- Osservatorio Astronomico di Capodimonte, via Moiariello 16, 80131 Napoli, Italy
        \and
        Jodrell Bank Centre for Astrophysics, School of Physics and Astronomy, University of Manchester, Oxford Road, Manchester M13 9PL, UK
        \and
        INAF -- Osservatorio Astrofisico di Arcetri, Largo Enrico Fermi 5, I-50125 Firenze, Italy
        \and Netherlands Institute for Radio Astronomy (ASTRON), Oude Hoogeveensedijk 4, 7991 PD Dwingeloo, The Netherlands
        \and
        Kapteyn Astronomical Institute, University of Groningen, PO Box 800, 9700 AV Groningen, The Netherlands
        \and
        Department of Physics, Informatics and Mathematics, University of Modena and Reggio Emilia, 41125 Modena, Italy
}

   \date{Received August 28, 2026; accepted October 23, 2026}

  \abstract{MAGNHIFFIC (MeerKAT AGN \ion{H}{I} Feeding \& Feedback Investigation close-by) leverages on sensitive MeerKAT 21-cm and VLT Survey Telescope (VST) observations of nearby Active Galactic Nuclei (AGN) to study feeding and feedback from the circum-nuclear ($\lesssim 1$ kpc) to the circum-galactic scales ($\gtrsim 10$ kpc). In nearby ($D\sim38$~Mpc) radio galaxy {\mbox NGC~3100}, these observations reveal, for the very first time, a starless filament of neutral atomic hydrogen ($M_{\rm H {\hskip 0.02cm \tt I}}\sim 3.4\times 10^8 {\rm M}_\odot$) extending from its outskirts ($\sim 340$ kpc) down to its innermost kiloparsecs. The kinematical properties of the filament show it likely formed by gas stripped from dwarf galaxy MKT095931-312323, flying by {\mbox NGC~3100}\ $\gtrsim 0.5$ Gyr ago. The filament appears to replenish the cold-gas reservoir of {\mbox NGC~3100}'s circum-nuclear disk traced by archival \element[][12]CO images as well as by \ion{H}{I} absorption. The kinematics of the \ion{H}{I} absorption and of the \element[][12]CO~(3--2) show that, in the very centre, the jets are currently stirring the multi-phase gas while dense clouds may still fall into the SMBH. The coexistence of a surviving cold reservoir, compact inner atomic/molecular clouds, and localised  jet-driven heating indicates that AGN feeding and feedback operate simultaneously rather than as separate evolutionary phases.}
  
	\titlerunning{\HI\ in \ntr}
	\authorrunning{F.~M. Maccagni, et al.}
\maketitle
\nolinenumbers

\section{Introduction}
\label{sec:intro}

This is the first paper of the MeerKAT AGN HI Feeding \& Feedback Investigation close-by (MAGNHIFFIC\footnote{\url{https://fil8.github.io/magnhiffic/}}) project, whose goal is to study feeding and feedback processes of jetted Active Galactic Nuclei (AGN) in nearby galaxies. These are multi-scale phenomena, where fuelling of the super-massive black hole (SMBH) and consequent ejection of energy into the surrounding medium occur and co-exist over three main spatial scales~\citep[\eg][]{Gaspari2020,Zajacek2022}. On the circum-nuclear scale ($0.1 - 1$ kpc), accretion onto the SMBH triggers the AGN and relativistic jets and/or radiative winds are injected into the interstellar medium (ISM). On the galactic scale ($1 - 10$ kpc), jets and winds can carve their way through the galactic body~\citep[\eg][]{Oosterloo:2017,Murthy2022,Audibert:2023}, while also bars, torques and turbulence may drive accretion onto the circum-nuclear scale~\citep[][]{combes:2013,combes:2014}. In the outskirts and environment of galaxies (circum-galactic scale, $10 - 100$ kpc) AGN jets can inflate gigantic lobes of relativistic particles~\citep[see, for example,][]{Fabian2012}, and galaxy interactions, ram pressure and condensation of the hot halo may favour gas to fall back into the galaxy~\citep[][]{Storchi:2019,Oosterloo2024}. Beyond spatial scales, temporal evolution is equally crucial; during the evolution of their host galaxies, SMBHs undergo multiple cycles of accretion and feedback, known as the AGN duty-cycle~\citep[\eg][]{Delvecchio2020,Brienza2025}. In detailed hydrodynamical simulations~\citep[\eg][]{Barbani2026a,Barbani2026b,Cammelli2026,Cammelli2026b,Piana2026a,Piana2026b}, this recursive loop can be described as a `weather' characterising the black hole environment, where the atmospheric state is regulated by the co-existence of feeding and feedback. At the onset of an AGN episode, the atmosphere may enter a {\em rainy} regime. Multiphase clouds and filaments seeded from environmental effects or condensing from the turbulent hot atmosphere stochastically rain toward the SMBH, fuelling the intermittent accretion more or less intensively while maintaining a cold-gas reservoir within the innermost kpc~\citep[][]{Gaspari:2013,Gaspari:2017}. 

Observationally, it is extremely difficult to characterise  the AGN weather over all scales. Optical Integral Field Unit instruments, and millimetre and sub-millimetre telescopes have a field of view limited to less than ten arcminutes. Even though they allow tracing the multiphase properties of AGN through both kinematical analysis and several diagnostics (e.g. ionisation rates, line ratios), most detailed studies of AGN feeding and feedback
are limited either to the circum-nuclear regions of nearby AGN~\citep[innermost kpc, \eg][]{Cresci2015,Mingozzi:2019}, or to particular regions along their jets and winds~\citep[][]{Santoro2016,Salome2016,Morganti2021}. 

Despite several detailed high resolution studies of AGN feeding and feedback mechanisms~\cite[\eg][]{Alatalo2015,Maccagni:2016,Dasyra2016,Maccagni:2018,Ruffa:2019a,Venturi:2021,Audibert2025,Zanchettin2025}, a comprehensive understanding of the AGN duty-cycle over all scales has yet to be obtained. It remains an open question how nuclear activity is fed and regulated over time, and how this leads to more or less efficient feedback across the galactic and circum-galactic scales. Furthermore, the specific mechanisms and timescales that self-sustain the AGN are still poorly understood~\citep[][]{Harrison:2018}. 

While all phases of the ISM (cold neutral atomic, molecular and hot-ionised) are involved in feeding and feedback loops~\citep[\eg][]{Oosterloo:2017,Maccagni:2021,Olivares2025}, most outflowing mass is cold~\citep[][]{Veilleux:2020}. The neutral atomic hydrogen seen at 21 cm (\HI) traces the bulk of the cold gas in galaxies. Its distribution and kinematics over the galactic and circum-galactic scales allows us to determine the environmental effects that trigger AGN and measure the timescales of these events. At circum-nuclear scales \HI\ absorption lines detected against AGN jets can trace gas inflows and outflows~\citep[\eg][]{morganti:2018}. However, most \HI\ clouds associated with AGN feeding and feedback phenomena are very diffuse, with column densities well below $\sim 10^{20}$~\cmsq. The lack of sensitivity of past radio interferometers limited these studies to either absorption features~\citep[\eg][]{vanGorkom1989,Gallimore:1999,Morganti:2005,Maccagni:2014,Maccagni:2017}, which are strongly hampered by the morphology of the underlying radio continuum, or to low spatial resolution ($\gtrsim 30$~\arcsec) observations of the AGN circum-galactic environment. 

A major step forward in the study of \HI\ in AGN was made possible by the Square Kilometre Array (SKA) precursor \meer~\citep[][]{Camilo:2018,Mauch:2020}. \meer\ has a unique combination of long baselines, enabling high resolution ($\lesssim 10$\arcsec) observations, a dense core of dishes, providing high sensitivity to diffuse gas ($10^{18-19}$ cm$^{-2}$), and a large field of view ($\sim 1.5$ deg$^2$). This makes \meer\ the one and only instrument today capable of detecting low-column density \HI\ in nearby AGN over all spatial scales. The uniform noise level of \meer\ over spatial scales of $6-90$ arcseconds allows us to detect low-column-density \HI\ emission ($10^{19}$~\cmsq) with kilo-pc resolution in nearby sources ($\lesssim 40$ Mpc)~\citep[][]{deBlok:2024,Maccagni2024b}. Indeed,~\citet{Maccagni:2026} demonstrates that deep \meer\ observations ($\sim 20$) open a new parameter space in \HI\ studies of AGN improving previous observations of one order of magnitude both in sensitivity and resolution and enabling detailed \HI\ studies on all relevant scales of the AGN feeding and feedback cycle~\citep[\eg][]{Maccagni:2020,Maccagni:2021,Santana2025,Mtshweni:2025}

Given the diverse characteristics of the AGN population, understanding the complexity of their feeding and feedback mechanisms requires detailed \HI\ studies of a representative sample of nearby galaxies. The MAGNHIFFIC project aims to undertake a systematic study of the \HI\ in nearby AGN with \meer. The MAGNHIFFIC's nearby AGN sample combines the Measuring Active Galactic Nuclei Under MUSE Microscope sample~\citep[MAGNUM, ][]{Cresci2015,Venturi2017,Mingozzi:2019,Venturi:2021} and the complete volume- and flux- limited radio galaxy sample extracted from the Parkes 2.7GHz survey~\citep[][]{Ekers:1989}, extensively with the Atacama Large Millimeter/sub-millimeter Array (ALMA) by~\citet[][]{Ruffa:2019a,Ruffa:2019b,Ruffa:2020,Ruffa:2022}. On the one hand, MAGNUM selected 10 Seyfert galaxies hosted by late-type star-forming galaxies (\eg\ Circinus, \mbox{NGC 4945}) or S0s (\ie\ Centaurus A). These galaxies are in clusters (\eg\ \mbox{NGC 1386} and \mbox{NGC 1365} in the Fornax Cluster), groups (\eg\ NGC 1068) and in more isolated environments (\eg\ NGC 5643). On the other hand, the radio galaxy sample covers AGN of low or intermediate radio power ($\log P_{\rm 1.4 GHz}\sim 23-25$~\whz) hosted by S0 or elliptical galaxies. Some of these AGN are the brightest galaxy of clusters or groups (\eg\ \mbox{NGC1399}, \forn). Together, these two dataset form a representative sample of low-to-intermediate power jetted AGN, which are numerically the dominant radio AGN population but whose impact on the evolution of their host galaxy and environment remains elusive~\citep[\eg][]{Girdhar2022}, and allow us to identify common trends and systematic differences in their feeding and feedback cycle. For example, most of the AGN energetic output of MAGNUM is radiative, but their low-power jets seem to generate a large turmoil in the ionised (and molecular) gas perpendicularly to their direction of expansions~\citep[][]{Venturi:2021}, while this effect is less evident in the radio galaxy sample, with \ntr\ being the only low-excitation radio galaxy where this may be occurring~\citep[][]{Ruffa:2022}. \HI\ observations trace the bulk of the cold gas reservoir over the galactic and circum-galactic scales and enable a direct comparison with the kinematics of the ionised and molecular phases over the circum-nuclear scales bridging the gap on the effects of feedback across all spatial scales. 

In this paper we present deep MeerKAT \HI\ observations of the first target of the MAGNHIFFIC sample: \ntr. These observations build on preliminary, shallower Australia Telescope Compact Array (ATCA) \HI\ observations~\citep[][]{Maccagni:2023}, and are analysed in combination with deep optical VLT Survey Telescope (VST) images, as well as archival ALMA data~\citep[][]{Ruffa:2019a,Ruffa:2019b,Ruffa:2020,Ruffa:2022}. In the following Section we describe the main properties of \ntr. In Sect.~\ref{sec:higroup}, we show the \HI\ detected in its group.  In Sect.~\ref{sec:hiAbs} we study in detail the kinematics of the \HI\ detected in absorption against \ntr's radio jets, while in Sect.~\ref{sec:hiFil} we focus on the distribution and kinematics of the `dark' \HI\ filament around \ntr. Section~\ref{sec:mol_res} presents a new central feature detected in the \cotre\ line and its possible link to the \HI\ absorption. In Sect.~\ref{sec:filFuel} we discuss the origin of the filament and its link to the fuelling of the AGN, while Sect.~\ref{sec:hi_co_centre} focuses on the nature of the \HI\ traced by the absorption complex and its relation with the \CO. Section~\ref{sec:jetism} presents the revisited $T_{\mathrm CO(2-1)}/T_{\mathrm CO (1-0)}$ line ratio map showing in detail the impact of the jets carving their path through the circum-nuclear disk. Our concluding remarks are in Section~\ref{sec:conclusions}.  \meer's 1.4 GHz observations achieve column density sensitivity at $20''$ over $10$~\kms\ of $2.7\times10^{19}$~\cmsq. We produce multi-resolution datacubes between $7$ and $90$ arcseconds following a similar observing strategy to the MHONGOOSE~\citep[][]{deBlok:2024} and MeerKAT Fornax Surveys~\citep[][]{Serra2023}. We refer to Appendix~\ref{sec:meerObs} and these works for a complete description of the \meer\ observations and data reduction. Appendix~\ref{sec:obsAlma} illustrates the main properties, data reduction and analysis steps of the VST and ALMA observations supporting our scientific analysis. We note that the archival ALMA \CO\ observations were reanalyzed at a spectral resolution of $\sim 2.4$~\kms\ comparable to the 1.4~\kms\ resolution of the \meer\ observations. Appendix~\ref{sec:atlas} shows the atlas of the optical and \HI\ data products of the \HI\ detected galaxies in \ntr's group.

\section{\ntr}
\label{sec:ntr}

\begin{figure*}  
 	\centering
 	\includegraphics[trim = 0 0 0 0, width=\textwidth]{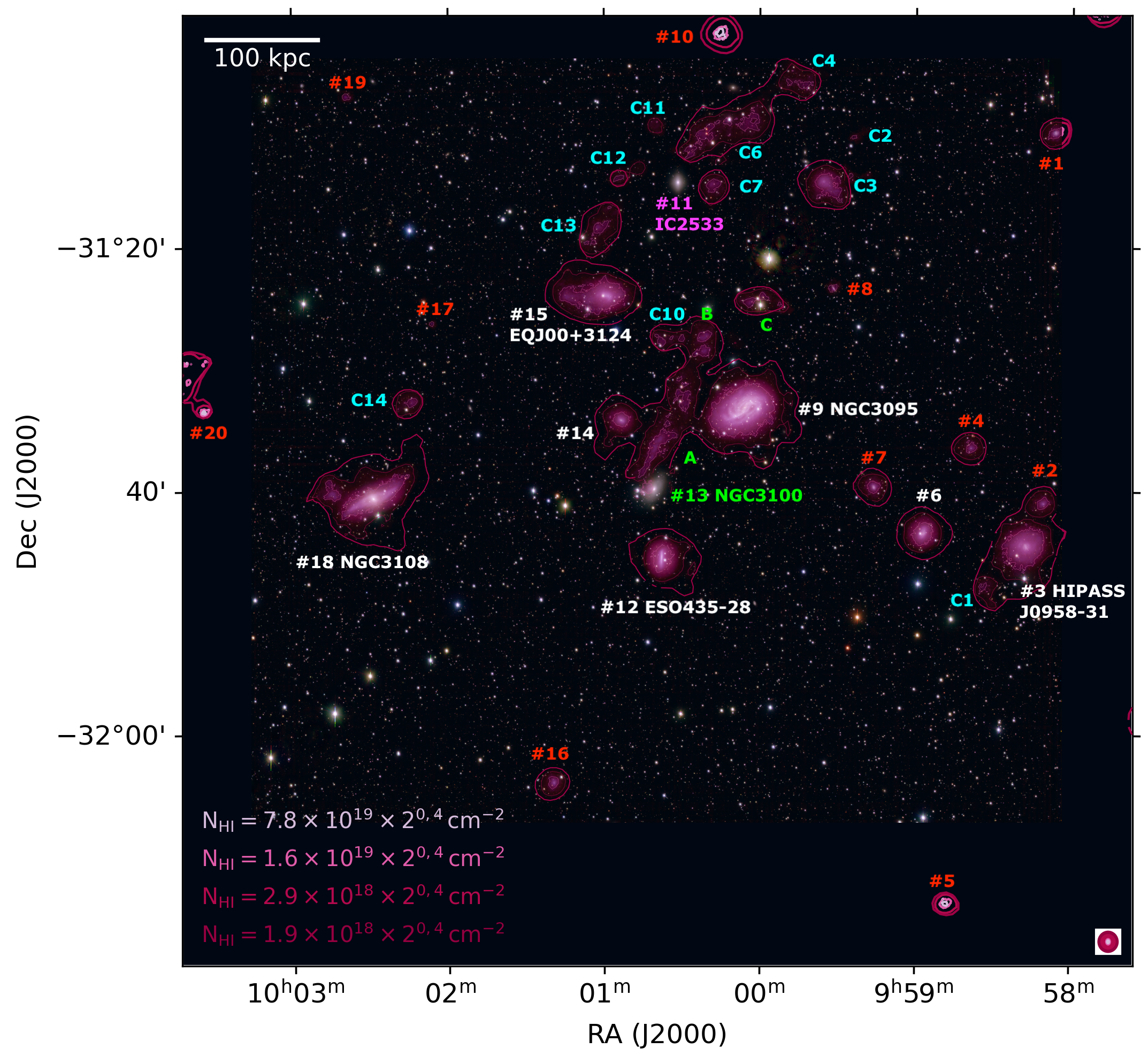}
    \caption{Optical and \HI\ emission in the NGC3100 group. The colour RGB image combines proprietary $i$-band VST observations with the VEGAS $r$- and $g$-bands from \citet{Iodice2021}. The shades of red from dark to light show the \HI\ emission from the $12$\arcsec$\times10$\arcsec, $32$\arcsec$\times23$\arcsec, and $65$\arcsec$\times64$\arcsec datacubes. The contours are the \nhi$=3\sigma$ contour for the same resolutions in addition to the $99\arcsec \times 94\arcsec$ datacube. Labels identify the source ID of the \HI\ detections (Table~\ref{tab:masses}). \HI\ sources detected by previous observations are labelled in white, while red and cyan labels mark novel \HI\ sources with and without optical counterpart, respectively. A, B, C label the `dark' filament, while ETG IC2533 (in magenta) has no associated \HI. The PSF of the  multi-resolution maps are in the bottom-right corner. Production of this RGB image followed the methods illustrated in~\citet{English2017}.}
 	\label{fig:fullFieldCont}%
 \end{figure*}

Among the MAGNHIFFIC sources, the early-type galaxy (ETG) \ntr\ (D=38 Mpc\footnote{Throughout this work we assume a standard $\Lambda$CDM cosmology with H$_{\rm 0}=70$\,km\,s$^{-1}$\,Mpc$^{\rm -1}$, $\Omega_{\rm \Lambda}=0.7$ and $\Omega_{\rm M}=0.3$. This gives a scale of $182$~pc/$''$, at the redshift of \ntr\ ($z=0.008$).}) is an ideal target where to study the complex AGN feeding and feedback cycle over all relevant scales because external gas supply from the circum-galactic environment, a circum-nuclear multiphase gas reservoir, and young jet perturbations are all visible in one system. Over the circum-nuclear scales, ALMA observations of three \CO\ rotational transitions~\citep[up to $J=3$,][]{Ruffa:2019a,Ruffa:2019b,Ruffa:2020,Ruffa:2022} reveal a circum-nuclear disk ($M_{H_2}=10^8$~\msun, $r\sim 1$ kpc, $i = 60^\circ, PA = 220^\circ$) seen, in projection, approximately perpendicular to the radio jets ($P_{1.4 \rm GHz} =1\times 10^{23}$~\whz, $t_{\rm AGN}\sim 2-5$~Myr). Kinematical modelling shows that the disk is overall regularly rotating ($v_{\rm rot} = 245$~\kms) but also that clouds with non-circular motions are present~\citep[][]{Ruffa:2020} of which some are associated to a jet-induced outflow~\citep[][]{Ruffa:2022}. Over the circum-galactic scales, deep ATCA $1.4$~GHz observations~\citep[][]{Maccagni:2023} show that six satellite galaxies have significant amounts of \HI\ gas distributed in a rotating disk with asymmetries and low-density tails. In particular, a diffuse \HI\ cloud is present between \ntr\ and nearby spiral galaxy \nzr. Deep optical {\em g,r} observations ($\mu_{g}\sim28.9$, $\mu_{r}\sim27.3$ mag arcsec$^2$) from the VST Early-type GAlaxy Survey~\citep[VEGAS, ][and references therein]{Iodice2021} do not show a stellar counterpart of the cloud. This starless cloud could be a remnant of a past interactions now forming a gas reservoir which may possibly replenish the centre of \ntr. Additionally, newly detected \HI\ absorption against the radio jets traces the neutral atomic counterpart of the molecular circum-nuclear disk. The low-resolution ($90$~\arcsec) ATCA observations do not allow us to study the kinematics of the `dark' cloud nor to understand the link (if any) between the \HI\ detected in absorption over the circum-nuclear scales with the `dark' cloud observed at the circum-galactic scales. How and over which timescales the AGN is fuelled remains an open question as well as how the jets impact the \HI\ observed in absorption at circum-nuclear scales. The deep \meer\ observations reveal the full elongated structure of the `dark' cloud, how it replenishes the circum-nuclear cold gas reservoir and its possible interplay with the radio jets.

\section{An \HI\ rich group of interacting galaxies}
\label{sec:higroup}

The deep ATCA observations discussed in~\citet[][]{Maccagni:2023} already showed new \HI\ features with two new group members detected ($\#6$ and $\#14$, see Table~\ref{tab:masses}). The \meer\ deep observations however are revolutionary  unveiling both low-mass, \HI-rich satellites surrounding massive group members (\eg\ NGC 3095, IC 2533, NGC 3108) and several \HI\ clouds between them, tracing intricate on-going interactions. \citet[][]{kourkchi:2017} report 18 galaxies associated to \ntr's group. With \meer, within 1 deg$^2$ and $500$~\kms\ from \ntr, we identify 13 new \HI-rich members and a plethora of \HI\ clouds. Figure~\ref{fig:fullFieldCont} shows the distribution of the \HI\ detected by \meer\ at resolutions between  $90$\arcsec to $7$\arcsec, overlapping with the $g,r$ optical images $i$-band from VST observations (see Appendix~\ref{sec:obsAlma} for details on the reduction and analysis of these images). Multiple \HI\ clouds are present around ETG \mbox{IC 2533}, possibly tracing the interactions with EQJ00+3124 (C11,12,13), and with low-surface brightness galaxy \#10 (C2,3, 4, 6, 7). \HI\ clouds also connect \mbox{EQJ00+3124} to \mbox{NGC 3095} (C10). In the West of \ntr, \mbox{HIPASSJ0958-31} is connected by a low-density \HI\ bridge to low-surface brightness galaxy (\#2) and cloud C1, suggesting an on-going interaction. In the East of \ntr, cloud C14 and the low-surface brightness galaxy \#17 are found along the direction of the warp of dusty ETG \mbox{NGC 3108}. At the edge of the image, source \#20 shows an extended \HI\ tail. Figure~\ref{fig:fullFieldCont} also reveals hat the `dark' cloud ($A$) is part of a large filament connecting the very centre of \ntr\ to a low-surface brightness companion $340$-projected kpc away (\#8).  The \HI\ properties of the detections with an optical counterpart are summarised in Table~\ref{tab:masses}. The field imaged in the $i$-band image is smaller than in the $g$ and $r$ images. Hence we rely on the VEGAS images to measure the surface brightness profiles of the \HI\ detected galaxies. Table~\ref{tab:stars} reports the total magnitudes ($g$ and $r$), the colour ($g-r$) and the stellar masses for the HI detected galaxies. In Appendix~\ref{sec:atlas}, for each \HI\ detection with an optical counterpart, we show the optical image with \HI\ contours overlaid, the $g,r$ surface brightness profiles, the \HI\ integrated spectrum, velocity field and position-velocity diagram across the major axis.

A complete description of the \HI\ in \ntr's group and of the stellar properties of its gas-rich members is worth a stand-alone analysis. Here, we focus on the \HI-to-stellar-mass content of the group members to complete the analysis done with the ATCA observations~\citep[][]{Maccagni:2023}, we identify which galaxies are \HI\ depleted and further support the presence of environment-driven gas stripping within the group. 

\begin{table*}[tbh]
        \caption{Properties of the \HI\ detected in the galaxies of the \ntr's group.}
        \centering
        \label{tab:masses}
        \begin{tabularx}{\textwidth}{l X c c c c c c} 
                \hline\hline                                                         
                ID & Source Name	& Coordinates &\vsys (\HI)   & $w50$  & $w20$   & F$_{\rm HI}$   &\mhi$^\dagger$    \\
                	      &             & [J2000] & [\kms]  &  [\kms] & [\kms]  & [\Jykms]   & [$\times10^8$\msun]   \\
                \hline

                 1  &\mbox{MKT J095806-311034} & 09:58:06, -31:10:34  & 2930 & 69 & 86 & 0.65 & $2.2\,(\pm0.4)$\\
                 2  &\mbox{MKT J095811-314100} & 09:58:10, -31:41:00  & 2037 & 58 & 69 & 0.33 &  $1.1\,(\pm0.3)$\\
                 3  & HIPASS J0958-31    & 09:58:12, -31:46:36  & 2806 & 69 & 95 & 6.3 & $21\,(\pm2.0)$ \\
                 4  &\mbox{MKT J095839-313625} & 09:58:38, -31:36:25 & 2944 & 32 & 53 & 0.02 & $0.57\,(\pm0.07)$ \\ 
                 5  &\mbox{MKT J095847-321354}& 09:58:47, -32:13:54 & 2285 & 61 & 35 & 0.30 & $1.0\,(\pm0.1)$ \\ 
                 6  &\mbox{J095858-314328}    & 09:58:58, -31:43:28 & 2616 & 93 & 113& 2.08 & $8.6\,(\pm0.7)$ \\
                 7 &\mbox{MKT J095916-313943} & 09:59:16, -31:39:43 & 2587 & 77 & 94 & 0.24 & $0.81\,(\pm0.09)$\\
                 8 &\mbox{MKT J095931-312323} & 09:59:31, -31:23:23 & 2738 & 55 & 79 & 0.06 & $0.20\,(\pm0.02)$ \\
                 9 &\mbox{NGC~3095}	       & 10:00:05, -31:33:22 & 2706 	& 338 & 365 & 22  & $75\,(\pm8.0)$	\\
                 10 &\mbox{MKT J100015-310223} & 10:00:15, -31:02:23 & 2640 & 40 & 77 & 0.63 &  $2.1\,(\pm0.2)$ \\
                 11  & IC 2533$^{\star}$                 & 10:00:31, -31:14:41 & -    &  - &  - &   -    &   -\\
                 12 &\mbox{ESO~435-028}        & 10:00:37, -31:45:29 & 2979 	& 128 & 158 & 2.5 & $10\,(\pm8.0)$	\\
                 13 &\ntr$^\ddagger$		       & 10:00:40,  -31:39:52&	2594   & 	178  &	397	& -0.056 $^\ast$ & $\ast\ast$ \\
                 14 &\mbox{J100053-313413}     & 10:00:53,-31:34:13	 & 2655 	& 96& 110 & 0.56& $1.8\,(\pm0.2)$	\\     
                 15 &\mbox{EQ J100-3124}       & 10:01:00, -31:24:01 & 2820 & 61 & 85 & 4.35 & $15\,(\pm2.0)$ \\
                 16 &\mbox{MKT J100120-320412} & 10:01:19,-32:03:59  & 2175 & 27 & 51 & 0.24 & $0.8\,(\pm0.09)$ \\
                 17 &\mbox{MKT J100206-312619} & 10:02:06,-31:26:18  & 2716 & 18 & 29 & 0.02 & $0.071\,(\pm0.002)$  \\
                 18 &\mbox{NGC 3108}           & 10:02:29,-31:40:38  & 2626 & 57 & 59 & 6.88& $23\,(\pm10)$\\
                 19 &\mbox{MKT J100238-310739} & 10:02:39,-31:07:39  & 2181 & 28 & 42 & 0.11 & $0.38\,(\pm0.07)$\\
                 20 &\mbox{MTK J100333-313326} & 10:03:33,-31:33:29  & 2700 & 34 & 59 & 0.09 & $0.33\,(\pm0.06)$  \\

                \hline                           
        \end{tabularx}
        \tablefoot{The properties are extracted from the 20\arcsec datacube which has the best trade-off between resolution and sensitivity and all sources are detected in this datacube. ($\dagger$): errors on \mhi\ are computed as the root sum square of the statistical error (\mhi/(S/N)) and the error on the flux scale ($10\%$). ($\star$): this ETG is the second most massive galaxy of the group and even though it does not show associated \HI\ it is closely surrounded by $7$ \HI\ clouds (see Fig.~\ref{fig:fullFieldCont}). ($\ddagger$): detected in absorption. ($^\ast$): corresponding to an optical depth of $\tau=0.114$. ($\ast\ast$): mass estimates from \HI\ absorption are given in Table~\ref{tab:absorption} and discussed in Sect.~\ref{sec:hiAbs}.}
\end{table*}

\begin{figure}  
 	\centering
 	\includegraphics[trim = 0 0 0 0, width=\columnwidth]{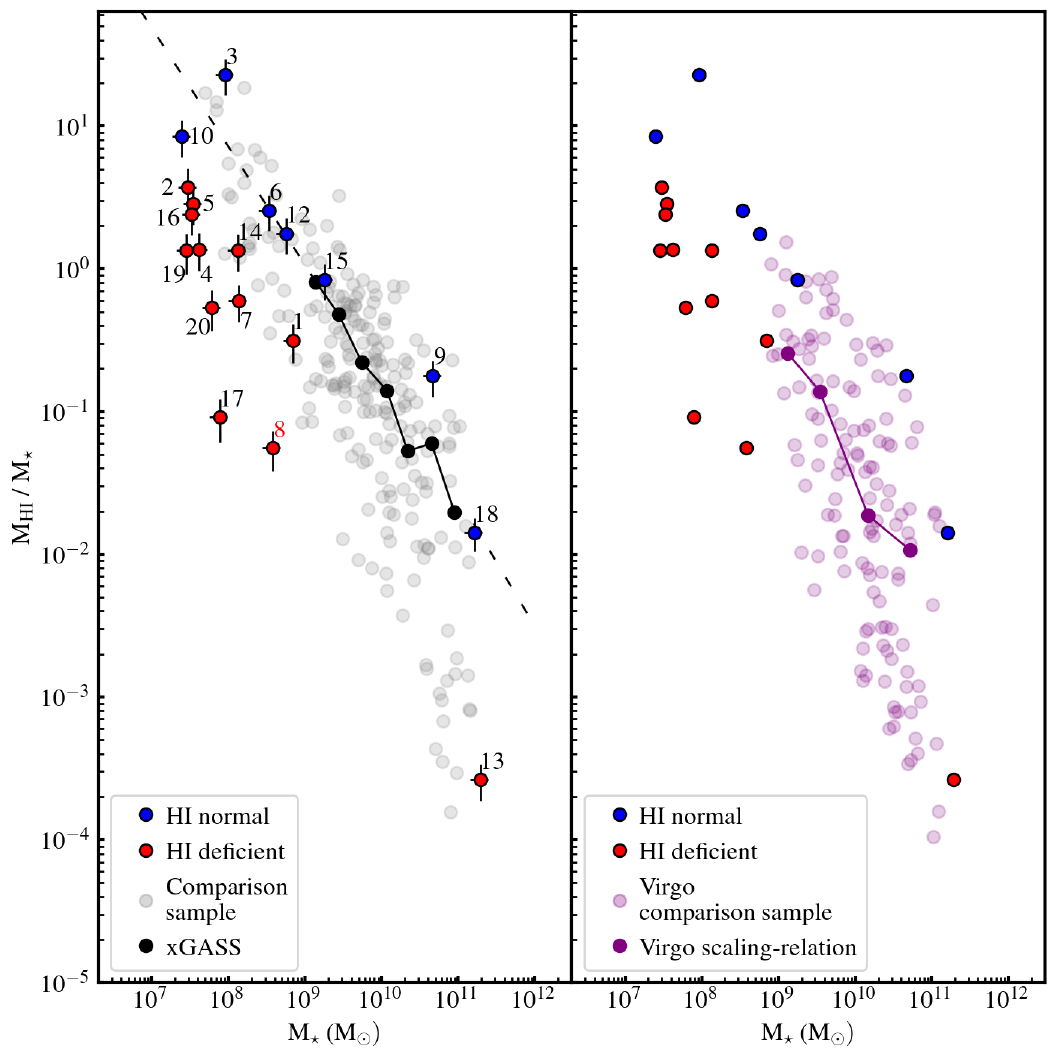}
	\caption{{\em Left Panel}: \HI\ fraction vs. the stellar mass of the galaxies in the \ntr\ group (red+blue markers) compared with non-cluster galaxies from VSG+HRS~\citep[grey circles, ][]{Kreckel:2012,Boselli2014}. Red and blue colours show \HI\ deficient and normal galaxies, respectively. The black solid line shows the xGASS scaling relation~\citep{Catinella:2018}. The black dashed line is the linear extrapolation of this trend for M$_\star~<~1.4~\times~10^9$~\msun\ and $M_\star~>1\times10^{11}$~\msun. {\em Right Panel}: Comparison between the \HI\ mass fraction of the galaxies in the \ntr\ group with that of Virgo cluster galaxies from HRS (light purple circles). Dark purple circles show the average scaling relation obtained from Virgo cluster galaxies~\citep{Cortese:2016}.}
 	\label{fig:hi_stellar_mass}%
 \end{figure}

Figure~\ref{fig:hi_stellar_mass} shows the \HI\ fraction (\ie\ the ratio between the \HI\ mass and the stellar mass) against the stellar mass of the galaxies detected with \meer. For comparison we show the behaviour of galaxies in poor environments of the Local Universe~\citep[left panel;][]{Kreckel:2012,Boselli2014} and in Virgo~\citep[right panel][]{Cortese:2016}. Compared to the same analysis done with ATCA~\citep[Fig.~6 in][]{Maccagni:2023}, now we are seeing several low-mass galaxies which appear \HI\ depleted, namely \#17, 8, 20, 19, 4, 7, 1, 5, 16, 5 and 14. \#17 is the most \HI\ depleted source, and it lies in the spatial and kinematical elongation of the warp of early-type dusty galaxy NGC3108 (\#18), confirming the scenario for which the \HI\ disk in this ETG originated from recent accretion from a companion~\citep[][]{Oosterloo:2002}. \#8\ is the second most \HI\ depleted galaxy and is found along the direction of the extended `dark' cloud (labels A, B, C).  All other \HI\ depleted sources are found at the edge of \meer's primary beam at low integrated S/N (between 5 and 7) and the statistical error on the mass is not negligible compared to the error on the flux scale (reported in Table~\ref{tab:masses} and Fig.~\ref{fig:hi_stellar_mass}). Galaxy \#13 is \ntr\ which has \HI\ detected only in absorption in the centre (see Sect.~\ref{sec:hiAbs}). Being an ETG is expected to have low \HI\ fraction, as the massive galaxies of the control samples. 

\section{The \HI\ absorption in \ntr}
\label{sec:hiAbs}

\begin{figure*}[tbh]
	\centering
	\includegraphics[width=\textwidth]{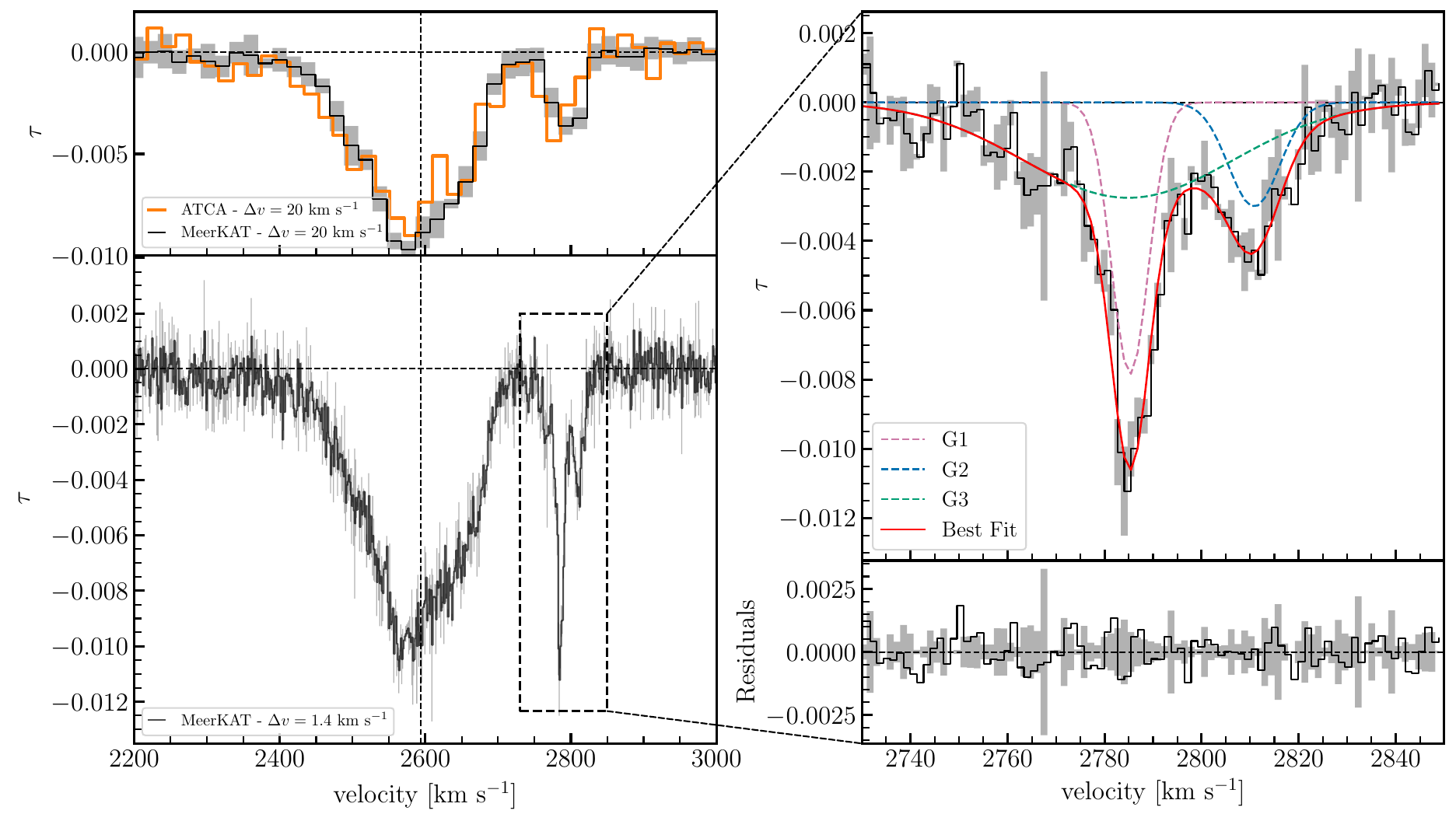}	
	\caption{{\em Bottom left panel:} \HI\ absorption detected in the centre of \ntr\ (with a PSF of 2\arcsec$\times$7\arcsec\ and 1.4~\kms\ spectral resolution) against the unresolved radio continuum emission of the AGN ($S_{\rm 1.4 GHz}\sim 600$ mJy), the grey shaded area shows the noise per channel, measured as the average r.m.s. within apertures of same size of the PSF. The dashed vertical line marks the systemic velocity of \ntr, \vsys$=2594$~\kms. {\em Top left panel:} Same spectrum compared with the one obtained with ATCA at the same 20~\kms\ resolution~\citep[orange line;][]{Maccagni:2023}. {\em Right panel:} zoom-in of the redshifted line complex. The red solid line shows the best fit and the red, blue and green dashed lines indicate its three Gaussian components.}
	\label{fig:absorption}%
\end{figure*}

\HI\ absorption was previously detected in \ntr\ by ATCA. Its line-shape and shift w.r.t.\ the systemic velocity of \ntr\ (\vsys$=2594$~\kms, determined from the 3D modelling of the \CO\ kinematics; \citealt{Ruffa:2019b}) allowed us to infer that the line traces the neutral atomic phase of the circum-nuclear disk of \ntr\ seen in emission in the CO~\cite[][]{Maccagni:2023}. Figure~\ref{fig:absorption} shows the absorption spectrum from the highest resolution datacube ($\sim 7$\arcsec, $\Delta v =1.4$~\kms). The top left panel of the Figure shows a comparison between the MeerKAT and the ATCA spectra both smoothed at the coarse resolution of the latter ($\Delta v=20$~\kms). The \meer\ observations fully recover the bulk of the absorption line tracing the disk and confirm the presence of a redshifted absorbing line at $v\sim2800$~\kms. Because of \meer's high spectral resolution ($\sim 1.4$~\kms), the redshifted component is resolved in a complex structure with two main peaks at $v=2784$~\kms\ and $v=2811$~\kms, respectively. At the ATCA resolution ($20$~\kms) the different features are spectrally unresolved and the redshifted complex has half the optical depth because of spectral smearing. The main properties of the \HI\ absorption lines are reported in Table~\ref{tab:absorption}.

From the \meer\ spectrum, we measure an integrated column density of the bulk of the absorption of $249\times10^{16}\,T_{\rm spin}/c_f$~\cmsq, where $T_{\rm spin}$ is the spin temperature of the absorption line and $c_f$ the covering factor of the background continuum source. Since the absorption is not resolved, to estimate the mass traced by the line it is necessary to make some assumptions on the covering factor of the absorbed \HI. In \citet[][]{Maccagni:2023} we have already demonstrated that the bulk of the absorption traces approximately one eighth of the circum-nuclear disk, the part seen against the background radio jets (the zoom-in box in the right panel of Fig.~\ref{fig:filamentMom0} shows the \couno\ velocity field of the disk and the AGN jets seen at 5 GHz). Since this region extends for $\approx4$ arcsec$^2$, assuming that the gas is uniformly absorbed over the region ($c_f=1$) and that it has $T_{\rm spin}=100$~K~\citep[as commonly done in \HI\ absorption studies,~\eg][]{Yoon:2025} the \HI\ mass is $2.6\times10^5$~\msun. If the \HI\ disk had uniform density, constant 100~K temperature and same extent of the \CO\ disk, it would have total \HI\ mass equal to $2.1\times10^6$~\msun.

The \meer\ spectrum allows us to characterise  the absorption properties and measure the velocity dispersion and integrated flux of the redshifted absorption complex. We perform a multi-Gaussian fit using the Python library {\tt lmfit}~\citep[][]{Newville:2016}. We fit up to ten Gaussian profiles and we discriminate the best fit based on the lowest BIC parameter. In the fitting procedure, we fix the minimum FWHM of the Gaussians to be equal to the channel-width of the spectrum ($1.4$~\kms) and the minimum height to be higher than the noise of the spectrum. The best fit model is shown in the right panel of Fig.~\ref{fig:absorption}. It results in three Gaussians: two of them (G1 and G3) are centred on the narrow and deep peak of the complex ($v\sim2785$~\kms), while the third (G2) is centred on its most-redshifted peak ($v\sim2811$~\kms). The main parameters of the Gaussians are shown in Table~\ref{tab:absorption}. The two narrow components (G1 and G2) have widths close to the typical velocity dispersion of the Cold Neutral Medium (CNM), indicating that the absorption traces compact dense gaseous clouds. The presence of a broad component (G3, $w_{50}=51.8$~\kms), twice the velocity dispersion of the warm neutral medium (WNM), suggests also the presence of a diffuse, turbulent \HI\ component.

The integrated column density of the redshifted complex is $63\times10^{16}\,T_{\rm spin}/c_f$~\cmsq, which is one order of magnitude lower than the typical column density of \HI\ associated to circum-nuclear disks of AGN~\citep[~\eg][]{Gereb:2015,Maccagni:2017} and similar to the low-column density of single clouds detected within the circum-nuclear disk of nearby AGN~\citep[as in, for example, \mbox{NGC~315} and \mbox{PKS 1718-649},][]{Morganti2009,Maccagni:2014}. Since the redshifted lines are unresolved against the radio jets, their distance from the SMBH is uncertain, as their size and mass. Nevertheless, we can make the following consideration. First, the absorption traces a spatially compact complex extending less than 1.2 kpc, since otherwise would have been detectable in emission in the datacube at the corresponding angular resolution (20\arcsec). At the redshifted velocities of the lines, instead, no \HI\ emission is detected, and the end of the filament lies at opposite blueshifted velocities (see Sect.~\ref{sec:hiFil}). Second, the low-velocity dispersion of the lines is consistent with \HI\ structures detected in the Milky Way~\citep[][]{Murray2015} and in nearby SF galaxies~\citep[][]{deBlok2026}, which extend from a few pc to a few hundred pc. Given that the volume number density of \HI\ features decreases with radius (see Fig.~\ref{fig:filamentMom0}), it is extremely unlikely that two such compact clouds lie along the same line of sight several kpc away from \ntr. It is more likely that these two clouds lie within the circum-nuclear disk. There, since the absorption intercepts only its central part~\citep[see Fig.
\ref{fig:mom_alma} and][]{Maccagni:2023} they are likely within its innermost 300 pc. Since the total integrated flux of the redshifted complex is approximately one tenth of the bulk of the absorption ($0.16$~\Jykms), a fiducial scaling to its \HI\ mass (assuming constant 100~K temperature and unity covering factor) is one tenth of the mass of the disk traced by absorption, \mhi$=2.4\times10^4$~\msun.

\begin{table*}[tbh]
        \caption{Properties of the \HI\ absorption complex of \ntr.}
        \centering
        \label{tab:absorption}
        \begin{tabularx}{\textwidth}{X c c c c c c} 
                \hline\hline  
                 Component                  &$v_{\rm peak}$ & $w_{50}$ &  $\tau_{\rm peak}$ & $F_{\rm int}$& \nhi$_{\rm int}$ & \mhi$_{\rm int}^{\dag}$\\
                        	                & [\kms]        & [\kms]   &                    &    [\Jykms]  & [$\times 10^{16}\times\frac{T_{\rm spin}}{c_f}$~\cmsq]   & [$\times 10^{5}$~\msun] \\
                \hline
                \mbox{Circum-nuclear disk}  &  2564           & 161      &      0.011         & 1.01             & 249             & 2.6  \\
                \mbox{Redshifted complex}  &  2785           &  -       &      0.011         & 0.16         & 63.0             & 0.24  \\
                \hline
                \mbox{G1}                   &  2785           & 9.0      &      0.008         & 0.045            & 18.2            & -         \\
                \mbox{G2}                   &  2811           & 12.4     &      0.003         & 0.024            & 9.8             & -         \\
                \mbox{G3}                   &  2785           & 50.9     &      0.003         & 0.089            & 34.8            & -         \\                
                \hline
                         \end{tabularx}
        \tablefoot{G1, G2, G3 refer to the three Gaussian profiles needed to describe the redshifted complex. The absorption lines are extracted from the {\tt r00} datacube with PSF $7.9$\arcsec$\times6.8$\arcsec. $^{\dag}$: the masses traced by absorption are computed assuming that the background continuum extends for $2\times7$\arcsec, $T_{\rm spin}=100$~K and uniform covering factor (see Sect.~\ref{sec:hiAbs}).}
\end{table*}

\section{The \HI\ filament of \ntr}
\label{sec:hiFil}

\begin{figure*}[tbh]
	\centering
	\includegraphics[width=0.44\textwidth]{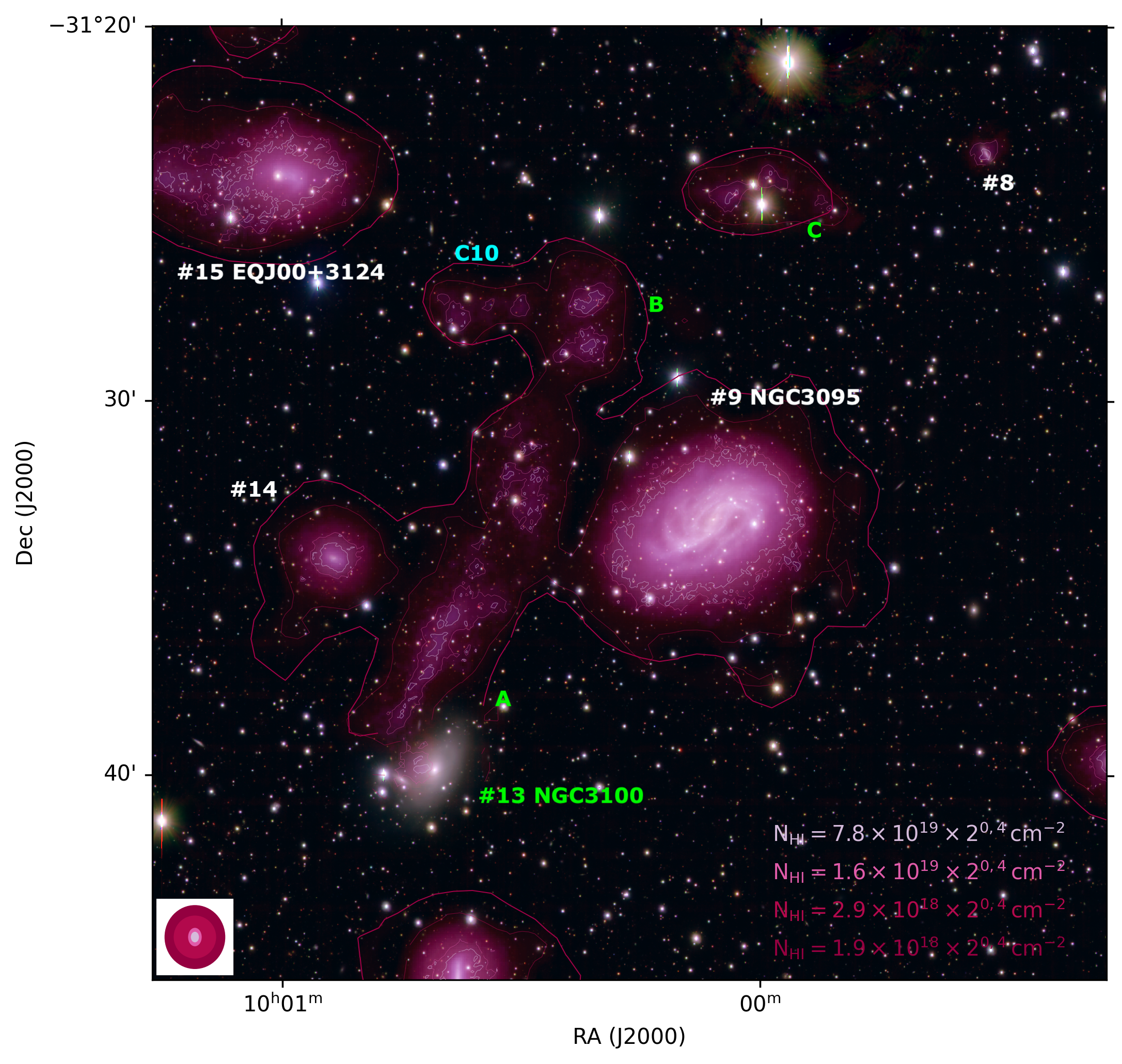}	
	\includegraphics[width=0.49\textwidth]{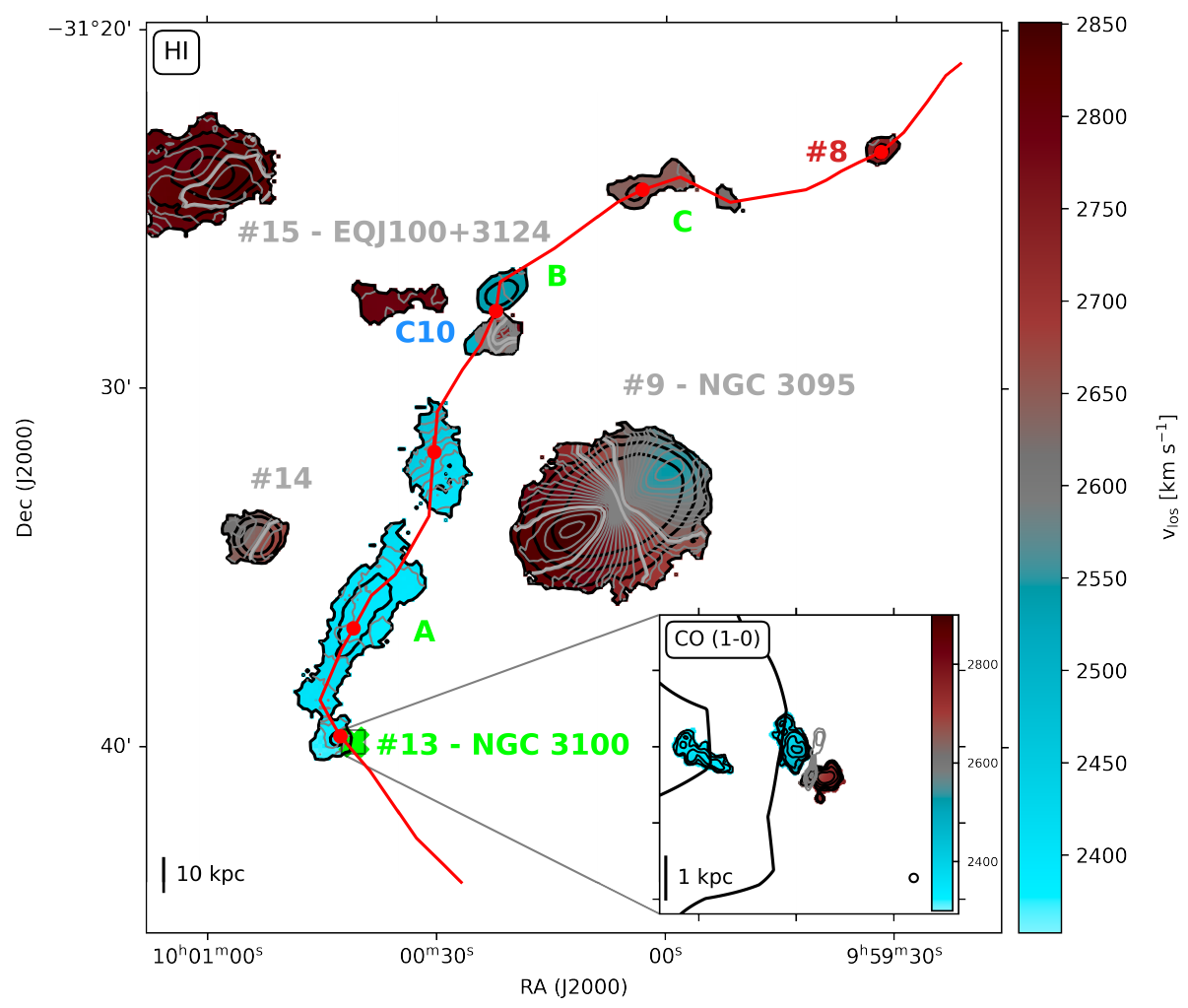}	
	\caption{{\em Left Panel:} zoom of Fig.~\ref{fig:fullFieldCont} into the \ntr, NGC 3095, EQJ00-3124 complex with overlaid the multi-resolution \HI. {\em Right Panel}: Velocity field of the \HI\ in the \ntr, NGC 3095 and EQJ00-3124 complex extracted from the $30$\arcsec datacube, with overlaid the flux density contours in black. The green cross marks the position of the centre of \ntr. The red line shows the path through which we extract the position velocity diagram in Fig.~\ref{fig:filamentPV}. The zoom-in box shows the velocity field of the \couno\ in the centre of \ntr\ and with its radio jets (see also Fig.~\ref{fig:mom_alma}), with the edge of \HI\ feature A in black contours (\nhi$=3\sigma\times2^n$ with $n = 0, 2$).}
	\label{fig:filamentMom0}%
\end{figure*}

The \meer\ observations show that the `dark' cloud is a filament extending from the centre of \ntr\ for $\approx340$ kpc in projected distance (Fig.~\ref{fig:filamentMom0}, left) and enable us to study its kinematics in detail. With {\tt iDaVIE}~\citep[][]{jarrett:2021,Sivitilli2026} we visually generate a 3D mask of the filament. With {\tt SoFiA-2}~\citep[][]{westmeier:2021} we identify that the filament is divided in three main components (A,B,C) whose \HI\ properties are summarised in Table~\ref{tab:hi-fil}. We report the measurements from the $30$\arcsec\ datacube (see Table~\ref{tab:cubes}) where we measure the highest total flux. At higher resolutions we measure lower fluxes, indicating that we are sensitive only to the clumpy component of the filament. The total mass of the filament is $3.4\times10^8$~\msun. The filament has 4 times more \HI\ gas than what measured with ATCA, which only detected part of A at 90\arcsec resolution. The left panel of Fig.~\ref{fig:filamentMom0} shows none of the clumps have a stellar counterpart, except the end of A which overlaps with \ntr. The velocity field (Fig.~\ref{fig:filamentMom0}, right) shows that the filament has a coherent velocity structure which is blueshifted and disconnected with the \HI\ of \#14, EQJ00+3124 and NGC3095. Part of feature B overlaps with C10 which has redshifted velocities and likely traces these latter two galaxies.

\begin{figure*}[tbh]
	\centering
	\includegraphics[width=\textwidth]{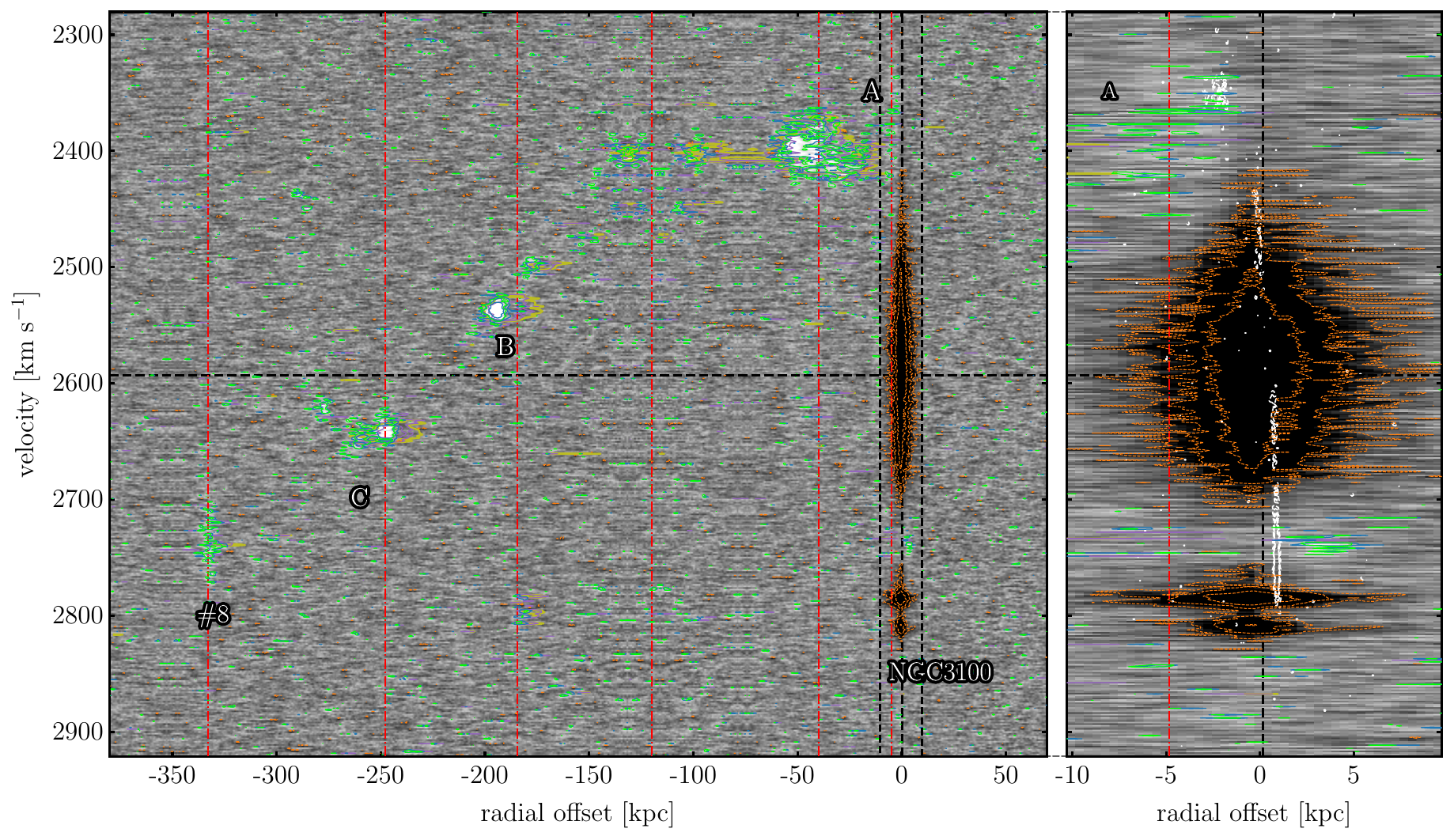}	
	\caption{Position-velocity diagram of the \HI\ filament of \ntr\ taken along the path shown in Fig.~\ref{fig:filamentMom0} (right panel) extracted the datacube at $30$\arcsec resolution. The filament has a coherent velocity structure from clump C at velocities $\sim 2750$~\kms\ in the north of NGC 3095 (see Fig.~\ref{fig:filamentMom0} to clump B and A spanning 300-kpc in projected distance down to the centre of \ntr. Cloud A is kinematically connected with the eastern \CO\ cloud (white contours, 2350~\kms) kinematically connected to the circum-nuclear disk traced both by \HI\ absorption (orange contours) and \couno\ emission. The Figure also shows \HI-depleted low-surface brightess galaxy \#8 after feature C, along the elongation of the smooth velocity gradient of the filament. Green contours mark the $ 3,5,10\times$ r.m.s of the slice, while orange contours show the negative fluxes of the absorption ($ 3,5,10\sigma$ contours, respectively). The horizontal black line shows the systemic velocity of \ntr\ (2594~\kms). The red vertical dashed lines show the offset of the points of the path shown the right panel of Fig.~\ref{fig:filamentMom0}. The black vertical dashed lines mark the zoom-in region of the centre shown in the right panel. The zoom-in right shows the \HI\ absorption, the \CO\ circum-nuclear disk (white contours) and the eastern \CO\ cloud .}
	\label{fig:filamentPV}%
\end{figure*}

The position velocity diagram along the filament (Fig.~\ref{fig:filamentPV}) shows that its components are connected by a smooth velocity gradient spanning from cloud A (showing a maximum flux at $v_{\rm peak}=2419$~\kms) to cloud C ($v_{\rm peak}=2664$~\kms). Along the prolongation of the gradient lies \HI\ depleted source \#8 ($v_{\rm sys}=2738$~\kms). This strongly suggests that the filament may trace the gas lost by this source moving through the group.   The zoom into \ntr\ (Fig.~\ref{fig:filamentPV}, right) also shows the \couno\ emission seen along the same cut in white contours. This traces the \CO\ circum-nuclear disk, visible within the \HI\ absorption, and the cloud in its east (at approximately 2350~\kms\ and 2 kpc of radial offset in the Figure). The latter has velocities coincident with the end of cloud A, indicating that the \HI\ filament ends into the eastern cloud of the \CO\ disk and is located within \ntr's innermost kiloparsecs and is not only seen in projection against it.

\section{The molecular gas in the centre of \ntr}
\label{sec:mol_res}

To further investigate the location of the \HI\ filament and of the redshifted absorption complex within \ntr, we compare the kinematics of the \HI\ with the ones of the molecular gas observed with ALMA in the \couno, \codue\ and \cotre\ transitions~\citep[][]{Ruffa:2019a,Ruffa:2019b,Ruffa:2022} reanalyzed at higher ($2.4$~\kms) spectral resolution (see Appendix~\ref{sec:obsAlma}).

\begin{figure*}[tbh]
	\centering
	\includegraphics[width=\textwidth]{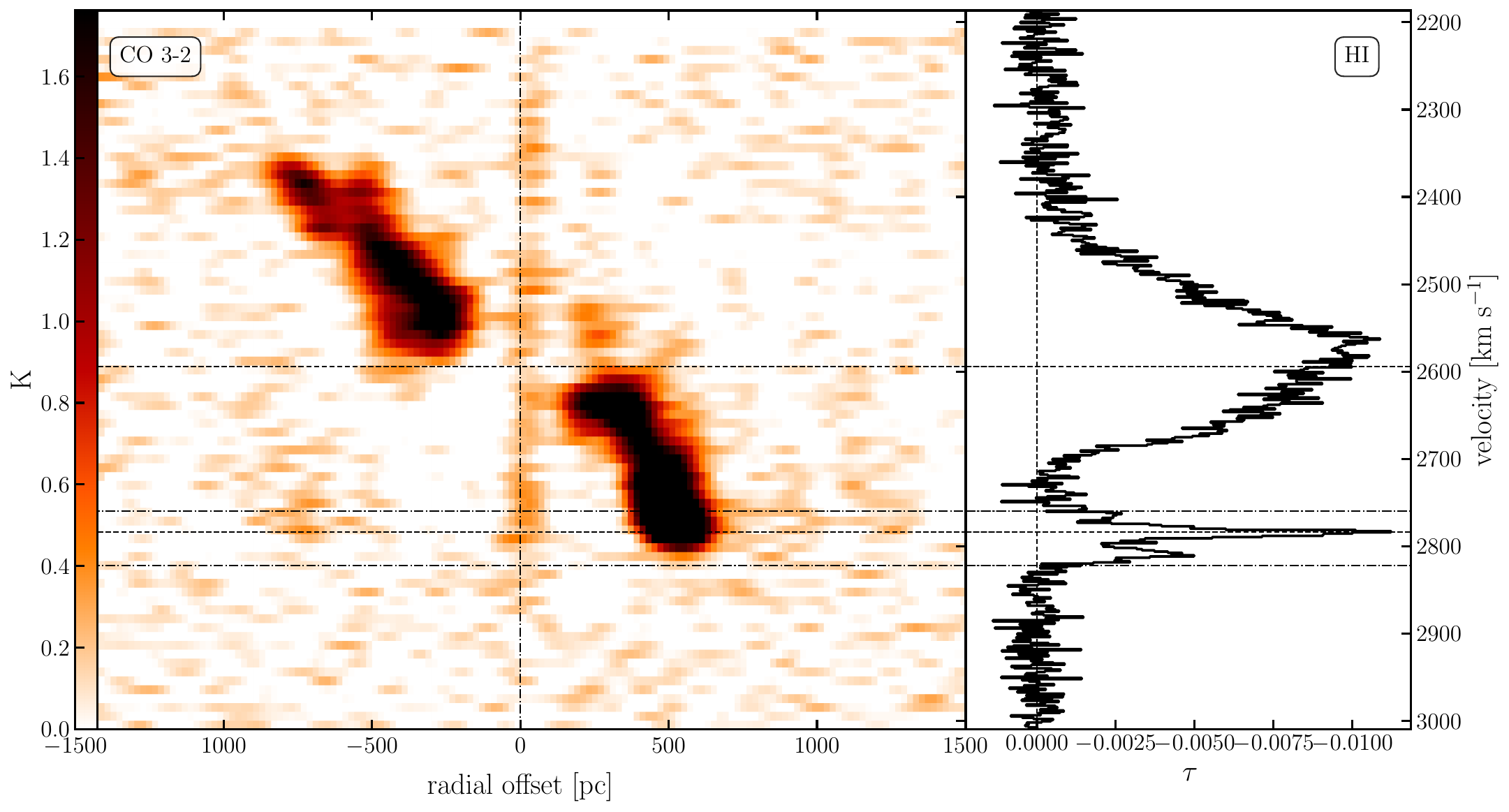}	
	\caption{{\em Left panel}: \cotre\ position-velocity diagram taken along the major axis of the disk ($PA=43^\circ$). To enhance the central feature we binned the datacube to $10$~\kms-wide channels. {\em Right panel}: \HI\ absorption over the same velocity range. In both panels, the upper horizontal dashed line marks the systemic velocity of \ntr\ ($2594$~\kms). The lower dashed line ($2784$~\kms) marks the peak of the narrow \HI\ line (G1), which matches the redshifted edge of the central \cotre\ feature, where it is brightest. The dash-dotted lines mark the velocity range of the redshifted \HI\ complex ($2760$--$2822$~\kms).}
	\label{fig:pv_cotre}%
\end{figure*}

We remark that in the centre of the disk, in the \cotre\ line is present a spectrally extended component which was unnoticed in the previous studies. This feature is clearly visible in the position-velocity diagram along the major axis of the disk ($PA=43^\circ$, Fig.~\ref{fig:pv_cotre}) extending between $2250$~\kms\ and $2850$~\kms\ with average $S/N=2.3$, integrated $S/N=13$~\footnote{The integrated $S/N$ ratio is computed by {\tt SoFiA-2} as $
\sum S_i/(\sqrt{N}\times\sigma)$, where $S_i$ is the sum of all flux densities of the feature, N the number of channels and $\sigma_{rms}$ the local noise.}. Only the highest peak of this feature ($S/N\gtrsim5$) is visible in the analysis of~\citet[][]{Ruffa:2022}, because the source finding was performed on the original resolution cubes. Using the smoothed cubes, it is possible to capture this unresolved, but spectrally extended, feature. This feature is seen only in the \cotre\ line while is not detected in the lower transitions. The fact that the feature is present in two datacubes independently of the self-calibration and continuum subtraction procedure used strongly suggests that this feature is not an artefact. Strikingly, the most redshifted velocities of this feature, where its emission is brightest, coincide with the peak of the narrow line of the redshifted \HI\ absorption complex (G1, $v\sim2785$~\kms). This is shown in Fig.~\ref{fig:pv_cotre}, where the horizontal dashed line links the peak of the \HI\ line to the \cotre\ feature, and the dash-dotted lines mark the velocity range of the \HI\ complex. This may indicate that the redshifted complex traces the neutral atomic counterpart of this feature. Further information on its nature and its possible link with the \HI\ absorption is given in Sect.~\ref{sec:hi_co_centre}.

Figure~\ref{fig:pv_cotre} also shows \cotre\ gas extending from the central edges of the disk beyond systemic velocity, into the top right and bottom left quadrants. This gas is partially visible in the masked position velocity diagrams shown in~\citet{Ruffa:2022}. Because it lies in the `forbidden velocity' quadrants the kinematics of this gas cannot be simply explained by regular rotation. This provides model-independent evidence to the presence and location of the \cotre\ outflow suggested by the high residuals in the kinematical modelling of~\citet[][]{Ruffa:2022}. From the regions with the largest residuals of this modelling, \citet{Ruffa:2022} estimated an outflowing molecular gas mass of $5.6\times10^4$~\msun\ and an outflow rate $\dot{M}_{\rm out}\lesssim0.12$~\msunyr.

\section{Discussion}
\label{sec:disc}

The \meer\ observations connect the $340$-kpc-scale external \HI\ reservoir, the kpc-scale cold gas disk, and candidate infalling clouds in the vicinity of SMBH ($r\lesssim 300$ pc). In the following sections, we discuss how these observations provide a multi-scale view of the AGN black hole weather, from the outer gas reservoir to the immediate environment of the jets

\subsection{The \HI\ filament fuelling the circum-nuclear disk of \ntr}
\label{sec:filFuel}

The most striking feature of \ntr's group is the \HI\ filament extending in projection for 340 kpc from \ntr\ into the intra-group medium. Combining the ALMA and MeerKAT data, we demonstrate that the end of the filament lies in the innermost regions of \ntr, and is not only seen in projection against it (Sect.\ref{sec:hiFil}). The velocities of the filament are approximately constant from the centre out to $\sim100$-projected kpc, which is inconsistent with gas being stripped off the galaxy, where we would expect a smooth gradient. In addition the eastern \CO\ cloud in the centre of \ntr, kinematically connected to the end of feature A (Fig.~\ref{fig:filamentMom0}) has also associated dust with sub-solar metallicities~\citep[][]{Ruffa:2022},  which also indicates an `external' origin of this gas. Taken together, the spatial and kinematic continuity is consistent with the filament replenishing the central cold-gas reservoir. 

The origin of the filament was so far unclear. \citet{Maccagni:2023} argued that it could be stripped gas from galaxy \#14, since this was the most \HI\ depleted SF galaxy in the surroundings of \ntr. The \meer\ observations reveal that \#14 is not kinematically connected with the filament, as well as C10 which likely traces the interaction between \mbox{NGC~3095} and \#15 (Sect.~\ref{sec:hiFil}). The dark filament could still be the remnant of these complex interactions that now is re-settled into a coherent kinematical structure and falling onto \ntr. A simpler possibility is that the filament is the remnant of gas stripped off a different satellite, undetected in the ATCA observations. As pointed out in Sect.~\ref{sec:hiFil}, the position-velocity diagram along the filament shows that source \#8 lies on the extension of its smooth gradient of velocities, suggesting a link between them. Galaxy \#8 has low surface brightness and it is the second most \HI\ depleted galaxy of the group (see Fig.~\ref{fig:hi_stellar_mass}), and the other \HI\ depleted galaxies are not close to the filament. We investigate further the \HI\ content of galaxy \#8 comparing the measured \HI\ mass with the expected \HI\ mass of the galaxy given its optical diameter ($D_{25}$, defined at magnitude 25) and morphological type, as done, for example, for the AMIGA sample and in the Virgo cluster~\citep[][]{Jones:2018,Boselli:2009} and originally proposed by~\citet{Haynes:1984a}. As shown in the surface brightness profile (\ref{fig:atlas_num8}, panel f), galaxy $\#8$ has $D_{25}\sim9$ kpc and the expected \HI\ mass is \mhi$^{\rm exp}=8.4\times10^8$~\msun~\citep[using the coefficients summarised in][]{Boselli:2009}. Since the measured \HI\ mass is $2.0\times10^7$~\msun\ the galaxy has more than one order of magnitude less \HI: \HI$_{\rm def} = \log M_{\rm H {\hskip 0.02cm \tt I}}^{exp}- \log M_{\rm H {\hskip 0.02cm \tt I}}^{obs}=1.62$. If instead, we compare the expected \HI\ mass of the galaxy with the measured \HI\ mass of the galaxy plus the total mass of the filament (\mhi\ (\#8+filament) $\sim3.4\times10^8$~\msun) the \HI\ deficiency is reduced considerably, \HI$_{\rm def} = 0.37$.  The \HI\ surface brightness profile and cumulative mass fraction (see Fig.~\ref{fig:atlas_num8}, panels g and h) also suggest a past interaction of source \#8, since the profile is flattened and only $50\%$ of the mass is contained within $r _{\rm exp} (M_{\rm H {\hskip 0.02cm \tt I} \#8}) \sim 2.5$~kpc, the expected \HI\ radius at surface brightness of 1~\msun\ pc$^{-2}$~\footnote{which corresponds approximately to $\gtrsim 1.2\times10^{20}$~\cmsq}, estimated from the \HI\ mass size relation,~\citep[][]{Wang:2016,Wang:2025}. This, along with its kinematical connection with the \HI\ filament, strongly indicates that passing through the environment of \ntr's halo \#8 lost most of its \HI, forming the `dark' filament that is currently replenishing \ntr\ of cold gas. 

If the filament had a tidal origin, the tidal forces must have been sufficiently strong to strip the gas but not strong enough to perturb the stellar body out to its edges, since the filament has no associated stellar counterpart down to $\mu_r\sim29$ mag arcsec$^{-2}$. Hence, the tidal radius of the interaction must have been at most equal to the initial $r_{25}$ of the galaxy. Assuming a typical baryon fraction of $5\%$, the total mass of \#8 plus the filament is $7.2\times 10^9$~\msun, while the dynamical mass of \ntr\ is $\sim 3.8\times 10^{11}$~\msun. Their dynamical mass ratio is $1.8\times10^{-2}$. A tidal radius equal to $r_{25}$ is obtained for a distance between the galaxies of $\sim27$ kpc. This indicates that source \#8 could have flown by \ntr\ as close as 27 kpc without perturbing its stellar body while losing its \HI. However, tidal interactions alone cannot explain the flattened \HI\ surface brightness distribution of the galaxy. Assuming that initially source \#8 had an \HI\ mass of $3.4\times10^8$~\msun, this should have been contained within its \HI\ radius, $r _{\rm exp}$ (\mhi\ (\#8+filament)$\sim 5$ kpc), expected from the \HI\ mass-size relation. This is approximately equal to $r_{25}$, hence other forces must have significantly stripped the \HI\ gas within $r_{25}$ without significantly perturbing the stellar body. Possibly, hydrodynamical forces between the galaxy and the group medium were also significant and caused ram-pressure stripping of the \HI.  Fig.~\ref{fig:filamentPV} shows that feature A has constant velocities, at its end it has an abrupt change (at $\sim 140$ projected-kpc), and then it connects with feature B and C with a smooth gradient, which further suggests that, after the interaction with \ntr, \#8 lost part of its \HI\ because of ram-pressure with the intra group medium. It is probable that the primary phase of tidal and hydrodynamical stripping took place at feature A's abrupt velocity transition, at a projected distance between 80 and 140 kpc from \ntr. Farther out, where tidal forces are weaker, ram pressure continued to shape features B and C, whereas the \HI\ of feature A is currently falling onto \ntr.

In the fly-by formation scenario of the filament, we can constrain the timescales of when it formed and understand how it could have fuelled the circum-nuclear disk. Assuming source \#8 is travelling at a velocity similar to the group velocity dispersion ($\sigma_{\rm group}\sim211$~\kms), and that closest distance from \ntr\ was $110$ kpc of the fly-by with \ntr\ would have occurred $\sim 1$ Gyr ago. The assumed velocity is reasonable because the galaxy's speed w.r.t \ntr\ along the line of sight is +100~\kms\ and it would only require to add a velocity vector in the plane of the sky of similar strength. If instead the galaxy was travelling at twice the speed ($420$~\kms, as a $2\sigma$ outlier of the velocity distribution of the group) it would have taken $\sim 0.5$ Gyr. The free-fall time within an isothermal potential provides a lower limit on the timescale of infall, $t_{ff} \approx 1.35 (r/c_{s})=0.395$ Gyr (where $r=110$kpc and $c_s=\sqrt5/3\sigma_{\rm group}=272$~\kms). If the fly-by occurred 0.5-1 Gyr ago, the \HI\ would have had sufficient time to fall into the centre of \ntr\ and form the circum-nuclear disk, which has a relaxation time of $0.095$ Gyr~\citep[][]{Ruffa:2019b}. More recently~\citep[2-5 Myr ago][]{Ruffa:2022}, instabilities within the circum-nuclear disk possibly triggered the AGN. While most of the atomic gas is in the filament, most of the molecular gas is in the circum-nuclear disk suggesting that the \HI\ converts into the molecular phase in the denser inner environment. \ntr\ hosts an externally seeded reservoir that joins a multiphase transport cascade at smaller radii, rather than classical in-situ precipitation of the filament itself from the hot atmosphere. Observations of all phases of the gas along the filament would allow us to understand how cooling occurs, if it starts in \ntr's environment or if the \HI\ converts to molecular only once reaching the regions of higher density, pressure and shielding, \ie\ the eastern \CO\ cloud. 

\subsection{\HI\ clouds in the centre of \ntr}
\label{sec:hi_co_centre}

As illustrated in Sect.~\ref{sec:hiAbs}, the \HI\ absorption in \ntr\ traces its circum-nuclear disk ($r\sim 500$ pc) and two distinct clouds with redshifted velocities. The distance of these clouds is uncertain but they are likely located within the innermost 300 pc of the circum-nuclear disk (see Sect.\ref{sec:hiAbs}). Their redshifted velocities with respect to systemic suggest that these clouds are  falling towards the SMBH. Similar, cold redshifted gas clouds have been detected in a few AGN~\citep[NGC315, A1907, PKS 1718-649,][]{Morganti2009,Tremblay2016,Maccagni:2018}. The morphology and width of the \HI\ redshifted complex is remarkably similar in width and shape to the redshifted \CO\ lines detected in A1907 which trace chaotic cold accretion (CCA), where condensation and cloud-cloud collisions can cancel angular momentum and drive intermittent SMBH feeding~\citep[][]{Gaspari:2013,Gaspari:2017}. 

\begin{figure}[tbh]
	\centering
	\includegraphics[width=\columnwidth]{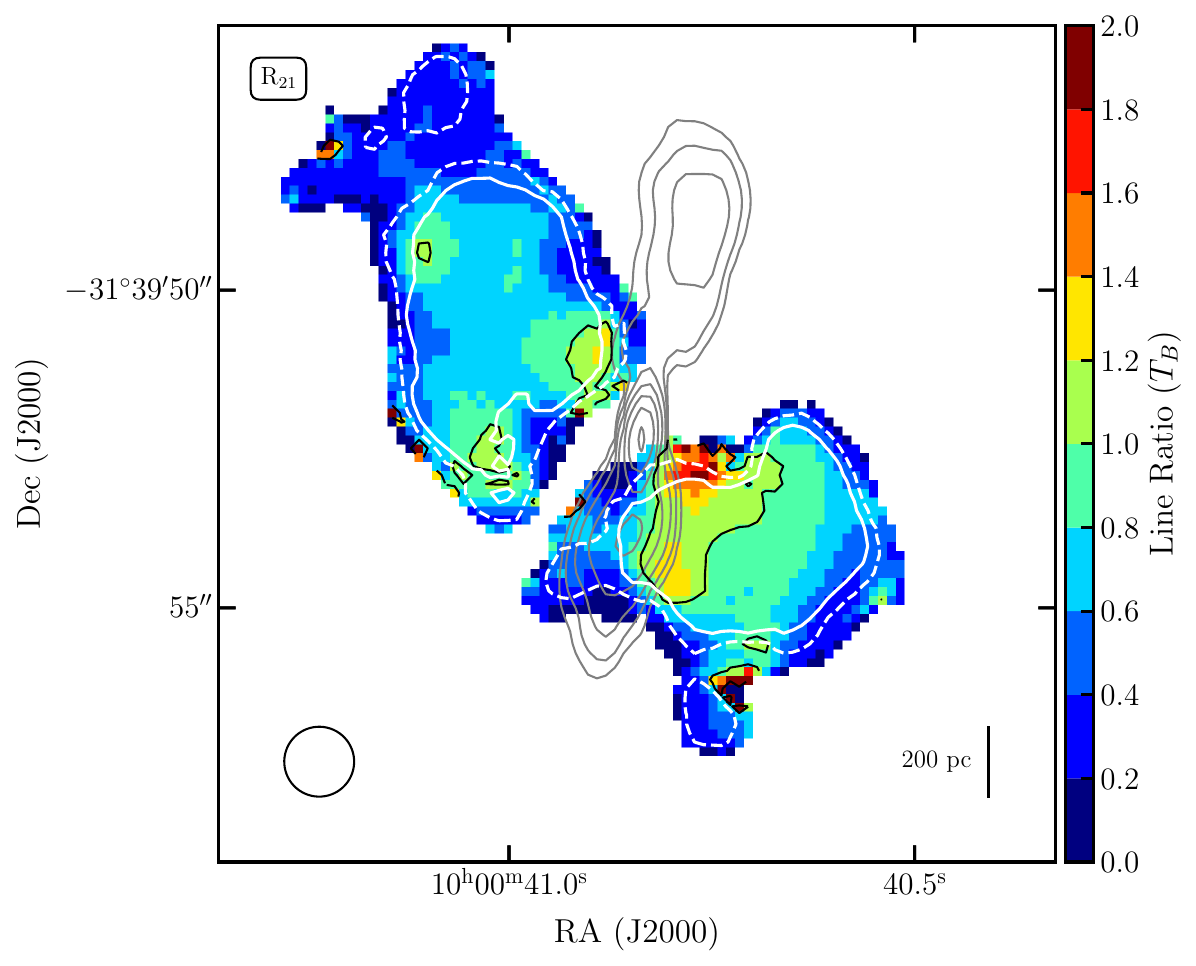}	
	\caption{$R_{21} = T_{CO(2-1)} /T_{CO(1-0)}$ map. The black contour shows the ratio of unity marking the transition between low-excitation gas and high-excitation gas. The white dashed and solid contours mark $R_{21}/\sigma_{R} = 3$ and $5$, respectively (see Fig.~\ref{fig:line_ratios_snr}). Overlaid on the image are the 5-GHz continuum contours with intervals of $1\times 2^n$~mJy (with n=0,1,2,...). The PSF ($1.1$\arcsec$\times1.1$\arcsec) is shown in the bottom left corner, while the corresponding scale length (200 pc) is in the bottom right corner.}
	\label{fig:line_ratios}%
\end{figure} 

In the innermost regions of \ntr\ the physical conditions differ from those in the circum-nuclear disk, since the \cotre\ central feature (see Fig.~\ref{fig:pv_cotre}) is not detected in the lower-J \CO\ transitions. This feature could trace a nuclear rotating disk. Considering the contribution of the SMBH ($M_{\rm SMBH} = 1.9^{+2.4}_{-1.1}\times10^8$~\msun, derived from the $M_{\rm SMBH}-\sigma$ relation for ETGs~\citep[][]{McConnell2013}) of the stellar mass distribution ($M_\star=1.9\times10^{11}$~\msun) and of the circum-nuclear disk~\citep[$v_{\rm rot}=245$~\kms, where kinematically modelling showed it warps with increasing radius from being face-on to edge-on][]{Ruffa:2022}, the velocity range of the feature can be explained only with an edge-on rotating disk in the innermost $20$ pc ($v_{\rm rot} \approx 200$~\kms). The slight extent at blueshifted velocities may indicate that even if dominated by rotation radial inflowing/outflowing motions are present (as also suggested by the residuals in the kinematical modelling of the \CO\ disk). Low-power Seyfert \mbox{NGC 1068} (also part of the MAGNHIFFIC sample) shows a similar cold-gas nuclear disk with streaming radial motions~\citep[][]{Rosas2025}. Interestingly, the peak of the narrow \HI\ redshifted line (G1, $v\sim2785$~\kms) has the same velocity as the redshifted edge of the central \cotre\ feature, where its emission is brightest. This kinematic coincidence could indicate that the \HI\ and the \cotre\ feature may trace the same cold gas in the innermost regions of \ntr. The \HI\ redshifted absorption could trace the edge of this nuclear disk, if only part of the disk lies in front of the radio continuum. If the disk were edge-on and entirely seen against a uniform background continuum, we would detect a double-horn absorption profile tracing its full rotation. If, instead, the background continuum is not uniform (as the structure of the unresolved jets) but intercepts on the line of sight only one edge of the disk, it would leave in the absorption profile only  one of the two horns. In \ntr, the approaching jet is expanding towards the south, where the circum-nuclear disk has redshifted velocities. If the nuclear disk shares the sense of rotation of the circum-nuclear disk, its southern side also has redshifted velocities, and the redshifted \HI\ complex could trace the part of the nuclear disk seen in absorption against the approaching jet. The \cotre\ feature could also trace a central bar, as suggested by the X-shape of the position-velocity diagram (Fig.~\ref{fig:pv_cotre}) considering both the feature and the disk and by the high residuals in the 3D kinematical modelling of the centre of the circum-nuclear disk~\citep[][]{Ruffa:2019b}. In this scenario, the redshifted \HI\ could trace the dense cloud inflowing within the bar. Alternatively, the \cotre\ feature could trace gas pushed radially outwards by the jets within the innermost 182 pc. With velocities comparable to the rotational velocity of the disk, this low-velocity outflow is unlikely to escape the central regions of \ntr. In this scenario the redshifted \HI\ cannot have velocities consistent with the \cotre\ feature but it would trace infalling clouds surviving the turbulent environment.  

Only sub-arcsecond resolution 1.4 GHz and ALMA observations could resolve the \HI\ absorption against the radio jets and determine its relation with the \cotre. Nevertheless, all presented scenarios suggest that the \HI\ redshifted absorption complex traces the atomic clouds surviving in the turbulent nuclear environment and likely falling onto the SMBH. 

\subsection{Jet-ISM interaction in \ntr\ and its black-hole weather}
\label{sec:jetism}

The turbulence caused by the radio jets onto the circum-nuclear disk is visible not only in the kinematical outflow (Sect.~\ref{sec:mol_res}) but also in \CO\ line ratios~\citep[][]{Ruffa:2022}. The high resolution datacubes allow us to trace this feedback effect in detail. Because of the complex velocity structure of the \cotre\ disk and since the out-flowing gas spatially overlaps with part of the West side of the disk (see Sect.~\ref{sec:mol_res}), integrated intensity ratios may obscure the distinct physical properties of overlapping gas clouds. We conservatively evaluate the \CO\ line ratios maps at high resolution dividing the moment-0 maps of the \CO\ distributions generated considering only the voxels of the datacubes where both lines are detected. Figure~\ref{fig:line_ratios} shows the $R_{21}$ line ratio map computed from the high-resolution datacubes (and Fig.~\ref{fig:line_ratios_32} the $R_{32}$ and $R_{31}$ maps). Regions with $R_{ij} \geq 1$ (black contour) show enhanced excitation and/or reduced optical depth~\citep[][]{Oosterloo:2017, Audibert:2023}. We also show the signal-to-noise maps of the line ratios ($R/\sigma_R$, Fig.~\ref{fig:line_ratios_snr}), whose 3 and 5 levels are overlaid as white contours in Figs.~\ref{fig:line_ratios} and~\ref{fig:line_ratios_32}. The highest $R_{21}$ ratios lie, in projection, along the western edge of the southern jet, and the bend of the southern jet is co-located with the \cotre\ outflowing gas. This may suggest that while expanding through the disk the jets may form a cocoon of shocked and heated gas around them, leaving most of the disk unperturbed, as predicted by the most recent simulations of jet expansion in a clumpy medium~\citep[\eg][]{Mukherjee2020,Meenakshi2022}.

The line ratios of \ntr\ are of the same order of magnitude as the ones measured in \forn~\citep[][]{Morokuma-Matsui2026}, while they appear less extreme than in IC5063~\citep[][]{Dasyra2016,Oosterloo:2017}, two other AGN of the MAGNHIFFIC sample which have been studied extensively in \HI\ and \CO. The jet-ISM coupling is most pronounced when the jet propagates along the major axis of the disk (\ie\ in IC5063) rather than when hitting only part of it as in \ntr\ and \forn. Hydrodynamic simulations of jet-ISM interactions confirm that the impact of jets on the surrounding medium strongly depends on their power and inclination with respect to the clumpy distribution of the circum-nuclear gas~\citep[\eg][]{Wagner2011,Wagner2012,Mukherjee2018}. 

The masses of the gas flowing in and out of the centre of \ntr\ are comparable: the \HI\ of the redshifted absorption complex, which may trace gas falling towards the SMBH, has mass $\sim2.4\times10^4$~\msun\ (see Sect.~\ref{sec:hiAbs}), while the jet-driven molecular outflow has mass  $\sim5.6\times10^4$~\msun\ ($\dot{M}_{\rm out}\lesssim0.12$~\msunyr;~\citealt{Ruffa:2022}). From the bolometric luminosity ($L_{\rm bol}\approx6\times10^{42}$~erg~s$^{-1}$) and the jet power ($P_{\rm jet}\approx10^{43}$~erg~s$^{-1}$) of the AGN~\citep{Ruffa:2022}, assuming a radiative efficiency of $0.1$, the accretion rate onto the SMBH is $\sim3\times10^{-3}$~\msunyr. Even for an accretion rate of $0.1$~\msunyr, over the lifetime of the radio source ($2-5$~Myr) the SMBH would have accreted only $2-5\times10^5$~\msun, less than $0.2\%$ of the \HI\ of the filament ($3.4\times10^8$~\msun). Possibly, at this stage of \ntr's nuclear activity, the jets perturb the circum-nuclear disk only locally, while the cold-gas reservoir continues to be replenished from the environment.   

MAGNHIFFIC observations of the \HI\ from the circum-nuclear to the circum-galactic scales in nearby AGN \ntr\ and \forn\ (in combination with ALMA, deep optical photometry and IFS observations) now enable us to start defining observationally the evolutionary scenario of the AGN duty-cycle, so far characterised in detail only in hydrodynamical simulations of galaxy groups atmospheres~\citep[\ie\ the stormy-sunny-rainy and cloudy weather loop,][]{Barbani2026a,Barbani2026b, Cammelli2026,Cammelli2026b,Piana2026a, Piana2026b}.

Both systems can be viewed as externally seeded system where the AGN {\em blackholeweather} is onset. In \ntr\ the $\sim340$-kpc \HI\ filament provides a macro-scale cold reservoir that was most likely acquired through environmental interactions rather than classical in-situ precipitation from the hot atmosphere. Once reaching the inner galaxy, this gas formed the circum-nuclear disk, where compact multiphase clouds may contribute to the fuelling process (\ie\ the redshifted absorption complex). In the last $\sim 5$ Myr, the young jets have locally heated and stirred the circum-nuclear disk, producing a cloudy-like inner state without globally destroying the cold phase. In \forn\ a gas-rich merger brought  $7\times10^8$~\msun\ of cold gas into its halo~\citep[][]{Serra:2019}. Since then, several episodes of nuclear activity generated the lobes as we see them now~\citep[][]{Maccagni:2020}. In the last 6 Myr, the radio jets expanded in the innermost 6 kilo-parsec unsettling and heating part of the gas, while, perpendicularly to them, a multi-phase CCA cascade re-triggered the core~\citep[][]{Maccagni:2021}. \ntr\ and \forn\ thus provide a rare observational bridge between environmental gas acquisition on halo scales and SMBH feeding on nuclear scales.

\section{Conclusions and final remarks}
\label{sec:conclusions}

Deep \meer\ 1.4~GHz observations of \ntr\ allow us to understand the feeding and feedback cycle of its AGN from the circum-nuclear to the circum-galactic scales. These observations reveal that in \ntr's group its three massive ETGs are interacting with \HI\ clouds and/or \HI-rich low-surface brightness galaxy. A dark \HI\ filament extends from the centre of \ntr\ out to $340$-projected kpc. The filament likely formed from the gas stripped off galaxy MKTJ095931-312323 (\#8, the most \HI-deficient galaxy kinematically associated with the filament), which likely flew-by \ntr\ between 0.5 and 1 Gyr ago, losing its \HI\ without strongly perturbing its optical body. Since then, the \HI\ fell onto \ntr\ and converted into molecular gas, as suggested by the kinematical connection of the filament with a \CO\ cloud in the innermost 2 kpc of the galaxy. Jointly with archival ALMA data, the \meer\ observations reveal that  feeding and feedback co-exist over the circum-nuclear scales. While the \cotre\ datacube shows an outflow ejected by the radio jets, the observed \HI\ redshifted absorption complex likely traces infalling clouds within $300$ pc surviving the expansion of the radio jets. The line ratios of the \CO\ show that the radio jets shock and heat the surrounding gas locally, forming a cocoon of excited gas while carving their path through the disk. These clues indicate that the jets are gently changing the conditions of the surrounding circum-nuclear regions (\ie\ feedback), while over the galactic and circum-galactic scales the infall (\ie\ feeding) is still on-going and compact cold clouds likely participate in a transport cascade towards the SMBH.  In \ntr\ the coexistence of a surviving cold reservoir, compact inner atomic/molecular clouds, and localised  jet-driven heating then suggests a cloudy-like circum-nuclear state in which feeding and feedback operate simultaneously rather than as separate evolutionary phases.

\begin{acknowledgements}
The authors thank the anonymous referee for the useful comments and suggestions. FMM carried out the research activities described in this paper with contribution of the Next Generation EU funds within the National Recovery and Resilience Plan (PNRR), Mission 4 - Education and Research, Component 2 - From Research to Business (M4C2), Investment Line 3.1 - Strengthening and creation of Research Infrastructures, Project IR0000034 – “STILES - Strengthening the Italian Leadership in ELT and SKA”.

JH acknowledges support from the UK SKA Regional Centre (UKSRC). The UKSRC is a collaboration between the University of Cambridge, University of Edinburgh, Durham University, University of Hertfordshire, University of Manchester, University College London, and the UKRI Science and Technology Facilities Council (STFC) Scientific Computing at RAL. The UKSRC is supported by funding from the UKRI STFC.

RR acknowledges financial support through INAF-WEAVE StePS funds and through PRIN-MIUR 2020SKSTHZ.498.

MG and FB acknowledge support from the ERC Consolidator Grant \textit{BlackHoleWeather} (101086804).

WJGdB and DK acknowledge funding from the European Research Council (ERC) under the European Union’s Horizon 2020 research and innovation programme (grant agreement No. 882793 “MeerGas”).

This paper makes use of MeerKAT data. The MeerKAT telescope is operated by the South African Radio Astronomy Observatory, which is a facility of the National Research Foundation, an agency of the Department of Science and Innovation. This paper uses data collected with the INAF VST telescope at the ESO Paranal Observatory.

(Part of) the data published here have been reduced using the CARACal pipeline, partially supported by ERC Starting grant number 679627 “FORNAX”, MAECI Grant Number ZA18GR02, DST-NRF Grant Number 113121 as part of the ISARP Joint Research Scheme, and BMBF project 05A17PC2 for D-MeerKAT. Information about CARACal can be obtained online under the URL: https://caracal.readthedocs.io.

\end{acknowledgements}

\bibliographystyle{aa} 
\bibliography{n3100HI.bib} 

\appendix 
\nolinenumbers
\section{Observations and data reduction}
\label{sec:obs}
\subsection{1.4 GHz \meer\ observations}
\label{sec:meerObs}

The 21-cm \meer\ observations of \ntr\ were taken in January and July 2023  in a project that targeted other two jetted AGN~\mbox{NGC3557} and \mbox{PKS 1718-649} (Proj. ID SCI-20220822-IP-01, PI Prandoni).  Each target was observed for 25 hours split in five separate 5 hour tracks. To maximise the uv-coverage of the targets, tracks are divided in two `rising', when the source is observed ascending towards zenith, and three `setting' tracks, when the target is observed while descending from zenith. Observations were taken at night to avoid solar interference which may contaminate the detection of diffuse low-column density emission.

\begin{table*}[tbh]
        \caption{Properties of the \magn\ \HI\ datacubes of \ntr}
        \centering
        \label{tab:cubes}
        \begin{tabularx}{\textwidth}{X c c c c c c } 
                \hline\hline                                                         
                Name        & robust&taper  & Beam              & Noise ($1\sigma)$ & $N_{\rm \HI}^{3\sigma, 10~\rm{km\,s}^{-1}}$&\mhi$^{3\sigma, 50\rm{km\,s}^{-1}}$\\
       	                    &       & ['']  & ['']              &   [\mJyb]         & [$10^{19}$~\cmsq]                           &[$10^7$~\msun]  	                  \\
       	        \hline
               {\tt r10\_t90}&$1.0$& $90$    & $99 \times 93$    & $0.50$              & $0.29$                                   &2.5                                         \\
               {\tt r05\_t60}&$0.5$& $60$    & $66 \times 65$    & $0.39$              & $0.50$                                   &1.9                                         \\
               {\tt r15}     &$1.5$& $0$     &  $35 \times 26$   & $0.25$	            & $1.5$                                   &1.2                                          \\ 
               {\tt r10}     &$1.0$& $0$     &  $26 \times 18$   & $0.24$	            & $2.7$                                   &1.2                                          \\ 
               {\tt r05}     &$0.5$& $0$     & $14 \times 10$    & $0.28$	            & $12$                                    &1.3                                          \\
               {\tt r00}     &$0.0$& $0$     & $7.9 \times 6.8$  & $0.35$	            & $35$                                    &1.7                                          \\
                \hline                           
        \end{tabularx}
        \tablefoot{(1) Name used to refer to the \HI\ datacube. (2) Robustness value. (3) Tapering used in arcseconds, 0 for no taper. (4) Average major and minor beam axis. (5) Average noise per channel. (6) $3\sigma, 10$~\kms\ column density sensitivity, which corresponds to approximately a S/N of 3 in the flux density map. (7) \HI\ mass sensitivity for a $3\sigma$, $50$~\kms\ unresolved source at the distance of the \ntr\ group.}
\end{table*}

At the beginning and at the end of each track, we observed a primary calibrator (either J1939–6342 or J0408–6545) for ten minutes to obtain time-independent flux, delay and bandpass solutions. Following the standard \meer\ observing strategy, every 30 minutes spent on the target, we executed a 2 minutes scan over the secondary calibrator (J1051-2023), which was used for the time-dependent frequency independent gain calibration. The data were collected both in the 'zoom' correlator mode (32 768 channels with width of 3.265 kHz) centred on the expected \HI\ frequency of the targets and in the `4k'-correlator mode (4000 channels throughout the \meer\ L-band, $\Delta\nu = 560-1712$ MHz and channel width of 208.984 kHz). Observations were taken with polarisation calibrator (3C138) for two scans of five minutes to exploit the `4k' not only for deep continuum studies but for polarisation as well. In what follows, we focus on the data reduction of the zoom-band observations used for \HI\ science. 

During extraction from the archive of the HH and VV polarisations we binned the data by two channels obtaining a measurement set of 5000 channels, each 6.53 kHz ($1.4$~\kms) wide for and approximate size of 0.5 TB. We processed the data using \cara~\citep{jozsa:2020}, a containerised pipeline developed in {\tt Python3} used for the reduction of \meer\ observations.  The entire process was carried out on a single node of a cluster, which features 128 cores and 1 TB of RAM. The same methodology and data reduction parameters have been used for all five tracks.

\ntr's observations achieve sensitivities and resolutions comparable to those of the MHONGOOSE survey~\citep[][]{deBlok:2024} and MeerKAT Fornax Survey~\citep[MFS,][]{Serra2023} and follow a similar observing strategy. Consequently, our data reduction process largely mirrors that of those surveys, with specific adjustments to account for the high radio brightness of the radio-loud \magn\ targets like \ntr\ ($S^{\rm peak}_{\rm 1.4 GHz}=492$~mJy). For a detailed explanation of each data processing step, we refer to ~\cite{deBlok:2024} and ~\cite{Serra2023}.

After flagging the calibrators and excluding the frequency range of the Milky Way's \HI\ emission, we derived the cross-calibration solutions by excluding baselines shorter than 100 m. Compared to the other \HI\ surveys mentioned above, we increased the bandpass smoothing to a 101 channel-wide running window. Similarly to MHONGOOSE and MFS, the self-calibration procedure was performed on the measurement set averaged in frequency to 1 MHz, the solutions were found using four spectral windows of approximately 8 MHz and fitting a second order polynomial between the frequency sub-bands. Using {\tt wsclean}~\citep{Offringa2014} and {\tt cubical} within the \cara\ framework, we performed three cycles of imaging and self-calibration. Cleaning was performed within a mask of the detected sources generated using {\tt SoFiA}~\citep[][]{Serra2015}. Unlike in MHONGOOSE and MFS, we performed self-calibration for both phases and delays with a time interval of 32s over the full sub-bands. 

Using {\tt crystalball} within \cara, we interpolated in frequency the clean model and self-calibrated gains. Besides the intensive memory usage requested by the interpolation in frequency, the continuum emission of \ntr\ requires up  to an order of magnitude more clean components with respect to the radio `quieter' fields of the Fornax Cluster or the MHONGOOSE star-forming galaxies. Continuum subtraction was done in three steps, first we subtracted the continuum model, second, in the {\em uv}-domain, we subtracted the model of the self-calibration solutions, third in the visibility domain by fitting a first order polynomial to the visibilities in the spectral range free of \HI\ sources.  

In interferometers residual broad-band RFI typical of the short baselines accumulates near the $u=0$ coordinates of the visibility domain~\citep[][]{Hess2015,Heald2016}, this may be due to poor bandpass filtering which cross-correlates when the fringes are zero. In the datacubes this RFI resembles broad horizontal stripes which would contaminate the detection of low-column density diffuse \HI. Before generating the \HI\ datacubes we executed two last flagging steps. First, We ran the flagging algorithm developed by~\citet[][]{maccagni:2022} to remove the `$u=0$' effect. Second, we applied an additional flagging to remove left-overs from previously flagged time-variable RFI~\citep[see ][for further details]{deBlok:2024}.

A single datacube cannot optimally capture simultaneously both the dense clumpy \HI\ SF of disks and the extended diffuse filament of the CGM. High-resolution ($\sim 10$\arcsec) uniform-weighted datacubes are ideal to study in detail \HI\ absorption spectra and \HI\ dense clouds, while low-resolution ($\gtrsim 30$~\arcsec) natural-weighted datacubes reach the sufficient combination of sensitivity and resolution to detect kpc-scale low-column density clouds. The latter typically trace environmental interactions as well as cold gas involved in the AGN feeding and feedback loop. For \ntr\ (similarly to MHONGOOSE and MFS observations) we produce six datacubes combining with different robustness and tapering to cover the entire range of angular resolutions, where MeerKAT's noise level is flat~\citep[$\sim 6$\arcsec - $90$\arcsec][]{deBlok:2024}. Table~\ref{tab:cubes} shows the main properties of the datacubes along with the $3\sigma$ column density sensitivity computed over $10$~\kms, which corresponds to a signal-to-noise ratio of three in the flux-density map, and the \HI\ mass $3\sigma$-limit computed over $50$~\kms\ at the distance of \ntr. 

On top of the multi-resolution datacubes, we also produce 2D maps of the flux-density, mean velocity and velocity dispersion distributions of the \HI, using {\tt SoFiA-2}~\citep[][]{westmeier:2021}. First we search for \HI\ detections with a smooth \&\ clip algorithm, with Gaussian kernels $1$, $2$ and $3$ times the spatial resolution of the cubes. The detection threshold is set to $3.5$ times the noise (evaluated in 3D over a spectral window of 7 channels). All \HI\ sources detected in the field of view have a detection reliability of $90$\%, based on the distribution of positive and negative noise peaks in the datacubes. We verify that {\tt SoFiA-2} identified all genuine \HI\ emission by also visually inspecting the datacubes in virtual reality~\citep[iDaVIE]{jarrett:2021}.

\subsection{Ancillary VST and ALMA observations}
\label{sec:obsAlma}

\begin{figure*}[tbh]
	\centering
	\includegraphics[width=\textwidth]{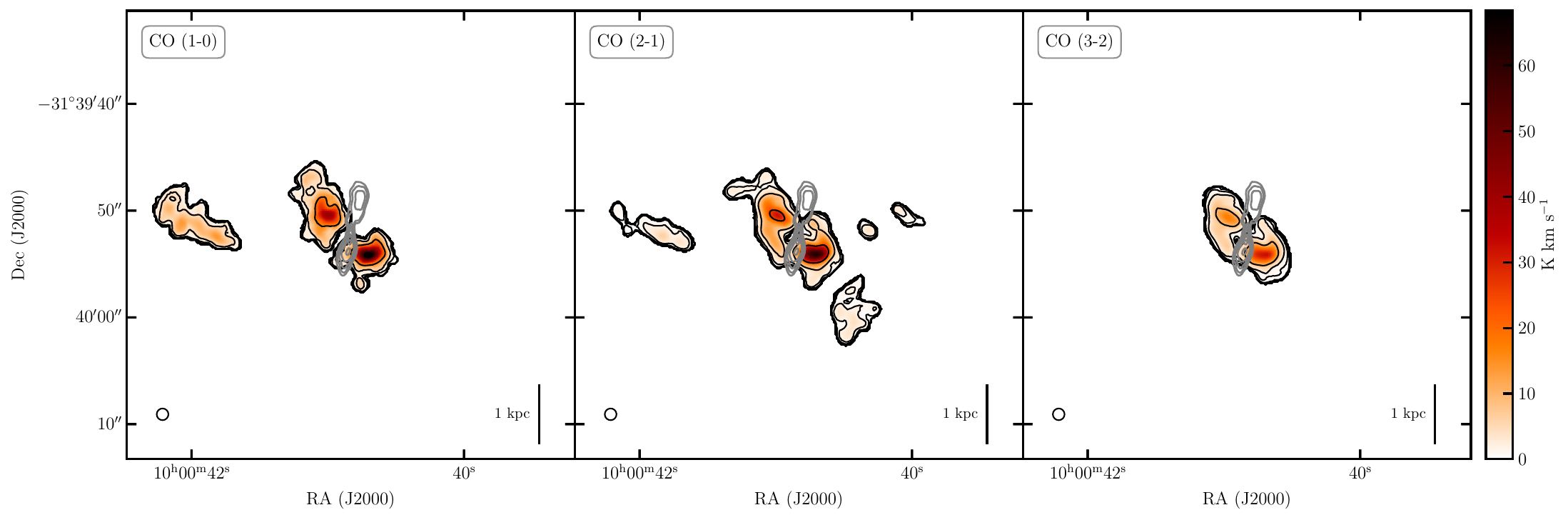}	
	\caption{{\em Left panel}: \couno integrated brightness temperature map. CO contour levels are $3\sigma\times2^n$ (n=0,1,2,..) and sigma is the noise over a single channel ($\Delta v=2.4$~\kms) is $1.6$~\mJyb. Overlaid on the image are the 5-GHz continuum contours with intervals of $1\times 2^n$~mJy (with n=0,1,2,...). The PSF ($1.1$\arcsec$\times1.1$\arcsec) is shown in the bottom left corner, while the corresponding scale length (200 pc) is in the bottom right corner. {\em Central panel}: same as in the left panel for the \codue\ transition, where the noise per 1.2~\kms-wide channel is $1.3$~\mJyb. {\em Right panel}: same as in the left panel for the \cotre\ line (the noise per 0.8~\kms-wide channel is $1.6$~\mJyb.}
	\label{fig:mom_alma}%
\end{figure*} 
In this paper we also make use of {\em g} and {\em r} photometric VLT Survey Telescope (VST) observations of \ntr\ taken as part of the VST Early-type GAlaxy Survey~\citep[VEGAS, ][]{Iodice2021}\footnote{\url{https://sites.google.com/inaf.it/vegas/home}} in run 0104.A-0072[B] and of the $i-band$ image acquired collected with the INAF VST telescope (B005-116.29GY, PI Maccagni). The wide-field of view ($\approx 1$ square degree) of VST perfectly matches \meer's, making its deep photometric images the best way to identify the optical counterpart of the \HI\ detections, including the smallest low-surface brightness galaxies~
\citep[][]{Maccagni:2024} and filaments of ionised gas~\citep[][]{Kleiner2021}. Thanks to the long integration time ($\sim 3.5$ and $\sim 2.75$ hours for the {\em g} and {\em r} filters, respectively), the VEGAS images reach the faint surface brightness levels of $\mu_g \sim 28.9$ mag arcsec$^{-2}$ and $\mu_r \sim 27.3$ mag arcsec$^{-2}$ in the g and r bands, respectively. The {\em r} and {\em g} images of \ntr\ were already used in~\citet[][]{Maccagni:2023} in combination with ATCA \HI\ observations, to which we refer for a complete description of the data reduction steps. Briefly, the images are reduced by using the dedicated AstroWISEpipeline~\citep[][]{McFarland2013}. As illustrated in~\citep[][]{Maccagni:2023}, the magnitudes of the galaxies are measured in the AB system and are corrected for Galactic extinction using the coefficients from~\citep[][]{Schlafly:2011}. We estimate the stellar masses following the procedure illustrated in~\citep[][]{Into:2013} using an average $M / L$ from the $g-r$ colours of the sources. Table~\ref{tab:stars} reports the integrated magnitudes, $g-r$ colors and stellar masses for all \HI\ detected galaxies in \ntr's group.

\begin{figure}[tbh]
	\centering
	\includegraphics[width=\columnwidth]{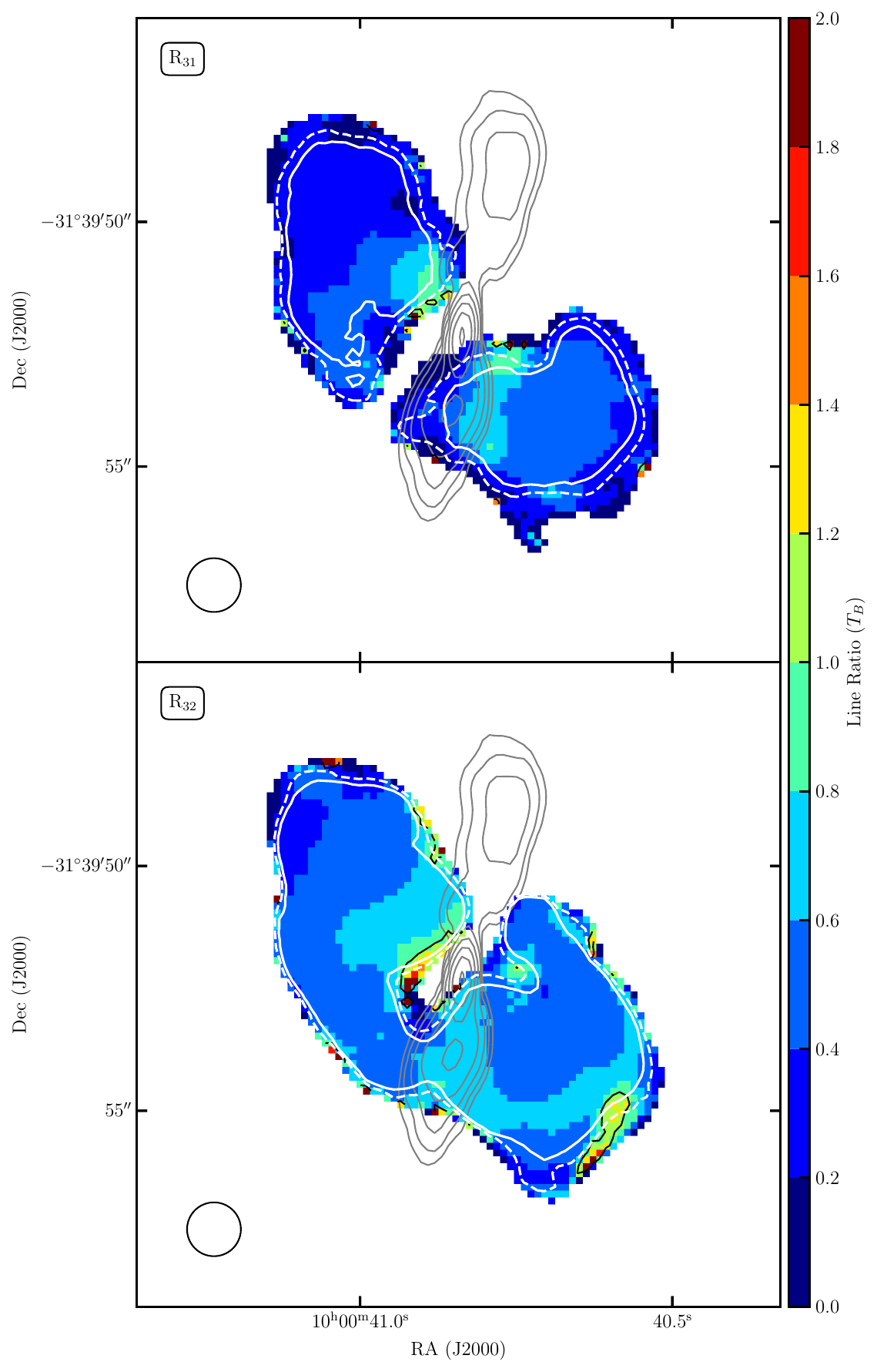}	
	\caption{{\em Top panel}: $R_{31} = T_{\rm CO(3-2)}/T_{\rm CO(1-0)}$ map. {\em Bottom panel}: $R_{32} = T_{\rm CO(3-2)}/T_{\rm CO(2-1)}$ map. Contours, beam and scale bar as in Fig.~\ref{fig:line_ratios}, with the white contours marking $R/\sigma_R = 3$ and $5$.}
	\label{fig:line_ratios_32}%
\end{figure} 

\begin{figure*}[tbh]
	\centering
	\includegraphics[width=\textwidth]{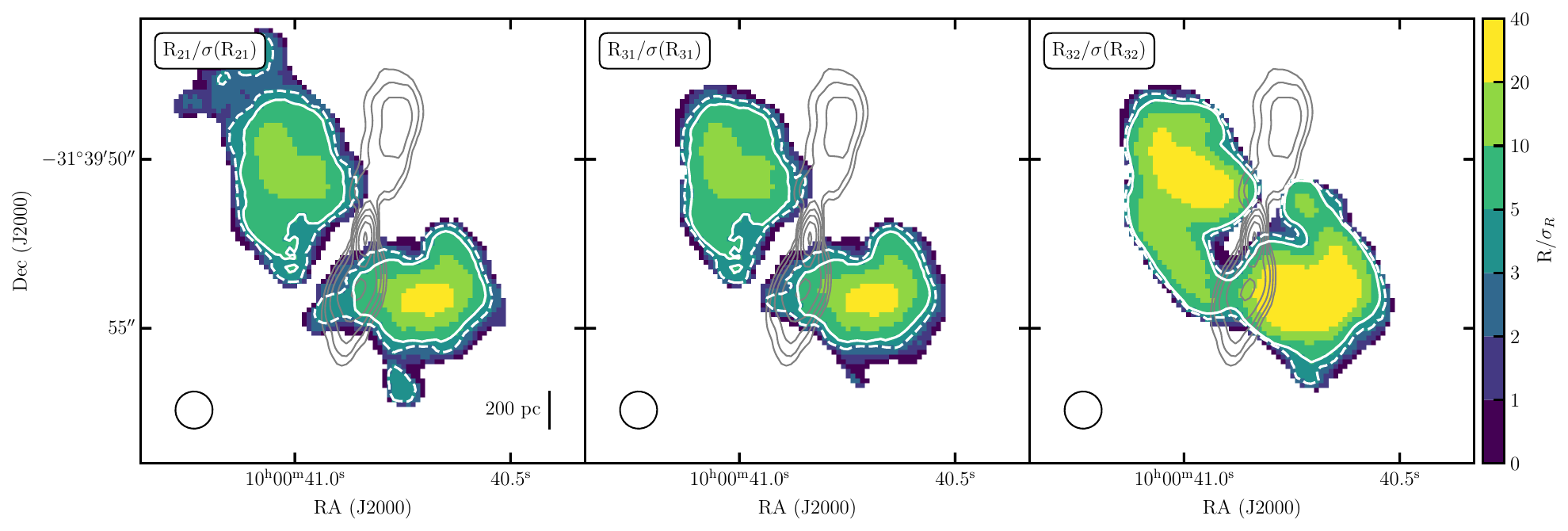}	
\caption{Signal-to-noise ratio, $R/\sigma_R$, of the $R_{21}$ (left), $R_{31}$ (centre) and $R_{32}$ (right) line-ratio maps. The white dashed and solid contours mark $R/\sigma_R = 3$ and $5$. The 5-GHz continuum contours, beam and scale bar are as in Fig.~\ref{fig:line_ratios}.}
	\label{fig:line_ratios_snr}%
\end{figure*}

To complete the study of the cold gas in the centre of \ntr, we combine the \HI\ observations with the ALMA high resolution observations ($\sim 1$\arcsec) of the \CO\ in the (1-0), (2-1) and (3-2) excitation lines. These observations were first presented in ~\cite{Ruffa:2019a,Ruffa:2019b,Ruffa:2022}, to which we refer for a complete description of their properties. These published observations were imaged at the coarse resolution of $10$~\kms. To best match \meer's spectral resolution, we analyse the original datacubes available on the ALMA archive ($\Delta v$ between $2.5$ and $0.8$~\kms, Proj IDs: 2015.1.01572.S and 2018.1.01095.S), rather than relying on the published $10$~\kms-wide channel datacubes~\citep[][]{Ruffa:2019a,Ruffa:2019b,Ruffa:2022}. These datacubes were produced automatically with the standard ALMA data reduction strategy and pipeline. For a uniform analysis of the three transitions, we regridded the datacubes to the same pixel scale and smoothed to a common resolution of 1.1\arcsec (except for the \cotre\ line in Fig.~\ref{fig:pv_cotre}) and $2.4$~\kms-wide channels. The r.m.s in the datacubes is $1.1$, $1.3$ and $1.7$~\mJyb\ respectively. After a visual inspection of the datacubes, we use {\tt Sofia-2} to reliably determine the regions where emission is detected and generate the flux density, velocity and velocity dispersion maps. Figure~\ref{fig:mom_alma} shows the flux density distribution of all transitions. In the new analysis we recover all features previously detected, in particular the cloud east of the disk in the \couno\ transition (see zoom in panel of Fig.~\ref{fig:filamentMom0}). The cloud has velocity range similar to the end of the \HI\ filament (see Sect.~\ref{sec:filFuel} for further details). We estimate the line ratio maps ($R_{21}\equiv T_{2-1}/T_{1-0}$, $R_{31}\equiv T_{3-2}/T_{1-0}$, $R_{32} \equiv T_{3-2}/T_{2-1}$, with $T$ in units of K~\kms) at high resolution dividing the moment-0 maps of the \CO\ generated considering only the 3D voxels where both lines are detected. Figure~\ref{fig:line_ratios_snr} shows the signal-to-noise of the line ratio maps computed propagating the noise maps of the single lines.

\begin{table}[tbh]
        \caption{Optical properties of the \HI\ detections in the \ntr\ group.}
        \centering
        \label{tab:stars}
        \begin{tabularx}{\columnwidth}{c X c c c c } 
                \hline\hline  
ID  & m$_g$ [mag]  & m$_r$ [mag] & $ g-r$ [mag] & $M_{\star}$[$\times10^{8}$\msun]   \\
\hline
1   &  15.45& 15.11    & 0.264  & $6.0$   \\
2    &  17.96& 17.81    & 0.150  & $0.29$   \\
3   &  17.51& 17.20    & 0.319  & $0.91$   \\
4   &  18.75& 18.36    & 0.366  & $0.36$   \\
5  &  18.07& 17.86    & 0.197  & $0.32$   \\
6 &  15.66& 15.44    & 0.220  & $3.4$   \\
7 &  16.07& 15.97    & 0.097  & $1.3$   \\
8  &  16.26& 15.89    & 0.369  & $3.3$   \\
9  &  11.81& 11.28    & 0.530  & $560$  \\    
10  &  17.48& 17.47    & 0.007  & $0.25$   \\
11 & 11.63 & 11.13    & 0.504  & $470$\\
12 & 15.53 & 15.22    & 0.310  & $6.3$   \\
13 & 11.46 & 10.68    & 0.780  & $1900$\\
14 & 17.00 & 16.71    & 0.290  & $1.5$   \\
15   & 16.17 & 15.47    & 0.700  & $19$   \\
16 & 18.12 & 17.91    & 0.214  & $0.33$   \\
17& 18.94 & 18.37    & 0.568  & $0.78$   \\
18 & 11.80 & 10.99    & 0.810  & $1500$\\ 
19& 19.31 & 18.89    & 0.420  & $0.28$   \\
20& 20.12 & 19.73    & 0.386  & $0.12$\\
\hline
         \end{tabularx}
        \tablefoot{The mean errors on the magnitudes are $0.05$ and $0.04$ mag for the $g$ and $r$ bands, respectively. The stellar mass is the average between the measurements in the two bands.}
\end{table}

\subsection{Atlas of \HI\ detected galaxies}
\label{sec:atlas}
In this Section we show the atlas of optical and \HI\ images of the \HI\ detected galaxies in \ntr's group. Table~\ref{tab:hi-fil} reports the \HI\ properties of the components of the \HI\ filament of galaxy \ntr\ (see Sect.~\ref{sec:hiFil}).

\begin{table}[tbh]
        \caption{Properties of the \HI\ filament of \ntr.}
        \centering
        \label{tab:hi-fil}
        \begin{tabularx}{\columnwidth}{X c c c} 
                \hline\hline  
                                            & A                 &B           & C\\
                \hline
                RA    [J2000]                      &  10:00:39		    &  10:00:22 &   10:00:04   \\
                Dec   [J2000]                     & -31:35:47         &-31:27:27  &-31:24:32\\
                $v_{\rm peak}$ [\kms] & 2419              & 2556      & 2664\\
                $w_{50}$[\kms]             &37                 & 19        &29\\
                $F_{\rm int}$ [\kms]        &0.684              &0.175&0.157\\
                \mhi$_{\rm int}^{\dag}$[\msun]&23               &5.8&5.2\\
                \hline
                         \end{tabularx}
\end{table}

\begin{figure*}[p]
	\centering
    {\includegraphics[width=0.55\textwidth]{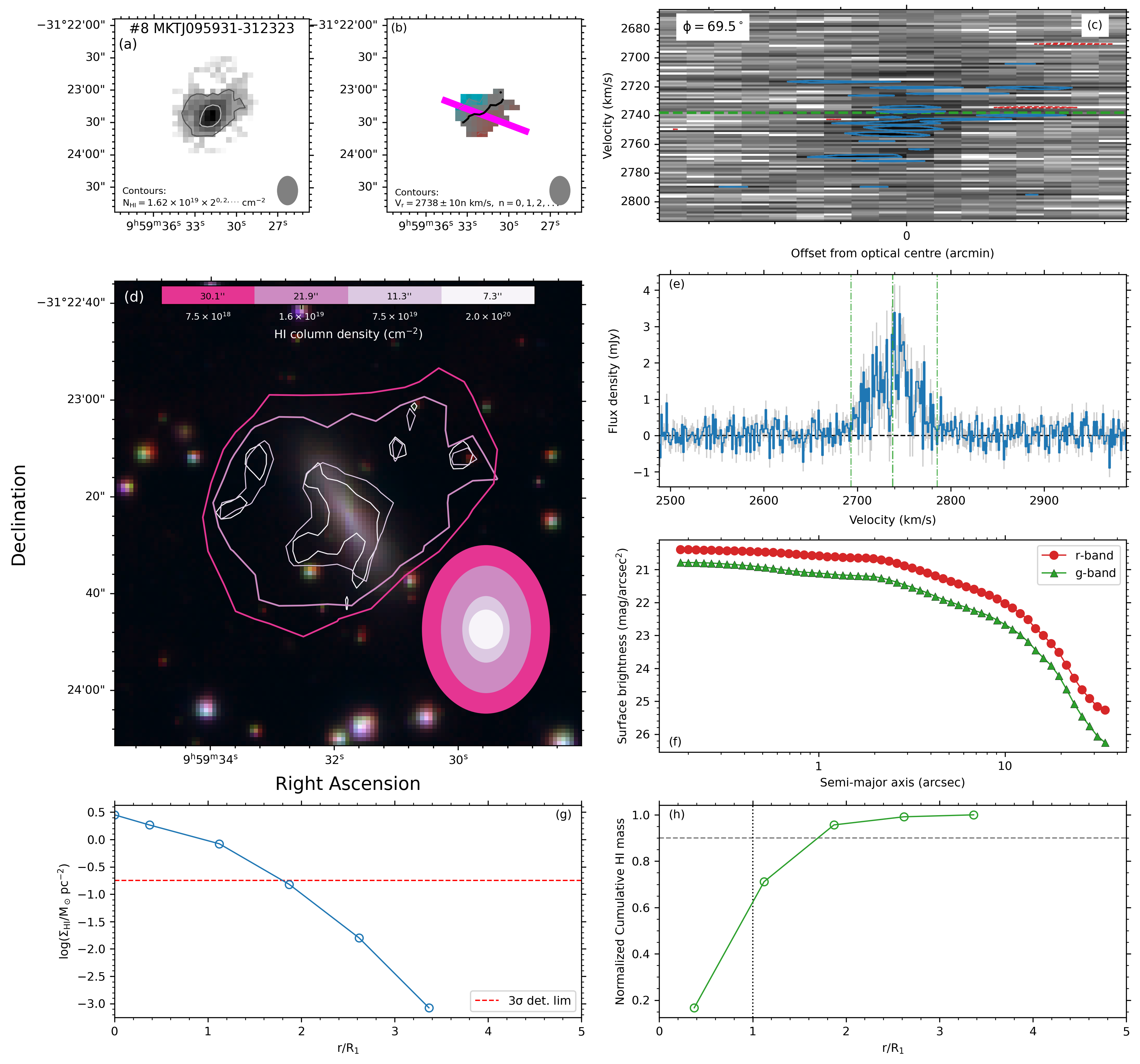}}
     \caption{{\em (a)} Total integrated intensity map of \ntr's filament progenitor, MKTJ095931-312323. {\em(b)} velocity field from the $32\arcsec \times 23\arcsec$ cube and {\em (c)} position velocity slice extracted across the major axis (magenta line). {\em (d)} RGB cutout from Fig.~\ref{fig:fullFieldCont} of target galaxy with N$_{\HI} = 3\sigma$ contour overlaid. {\em(e)} Integrated spectrum of target galaxy. {\em(f)} $r$- and $g$-band surface brightness profiles represented by the red circles and green triangles respectively. {\em (g)} \HI\ surface brightness profile normalised by the $1\,$\msun$\,\text{pc}^{-2}$ radius. {\em (h)} Normalised cumulative \HI\ mass as function of the $1\,$\msun$\,\text{pc}^{-2}$ radius.}
	\label{fig:atlas_num8}%
\end{figure*}

Figure~\ref{fig:atlas_num8} shows a multi-panel view of the \HI\ and optical properties of the candidate filament's progenitor, source $\#8$ MKT095931-31219. Panel (a) shows the \HI\ flux density map at $12$\arcsec, panel (b) shows the corresponding velocity field and panel (c) the pv-diagram across the major axis of the \HI\ disk. Panel (d) shows the combined $g,r,i$ colour image from VST. Panel (e) shows the integrated \HI\ spectrum. Panel (f) shows the $g,r$ surface brightness profiles measured in the VEGAS images. Panels (g) and (h) show the \HI\ surface brightness profiles and the cumulative \HI\ mass, respectively. Figure~\ref{fig:atlas} shows panels (a) to (f) for all \HI\ detected galaxies in \ntr's group. For the sources outside the field the VST $i-band$ image, we generate the $gri$ image from the DeCaLs data release.

\begin{figure*}[p]
	\centering
    \rotatebox{90}{\includegraphics[height=0.4\textwidth]{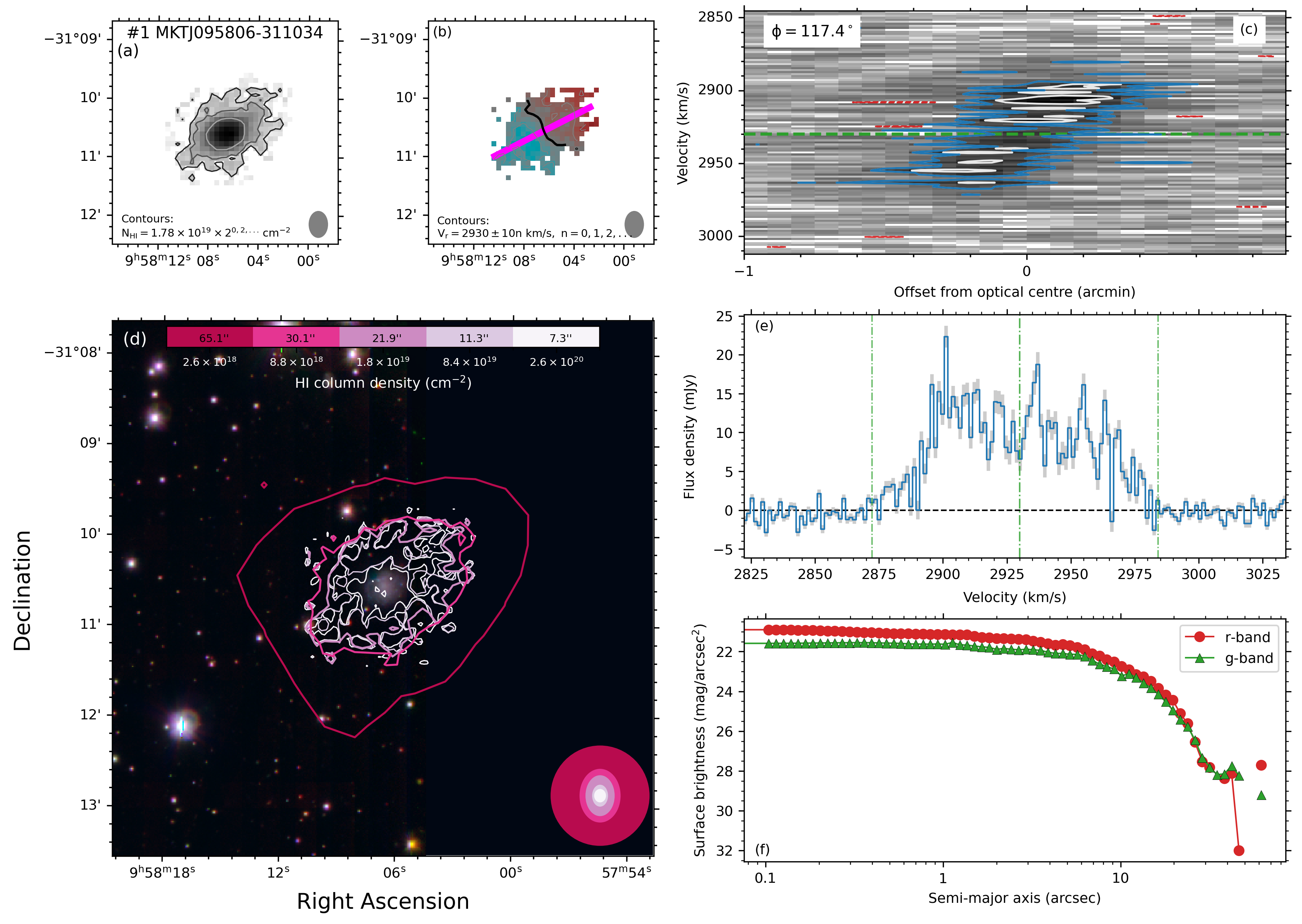}}
    \rotatebox{90}{\includegraphics[height=0.4\textwidth]{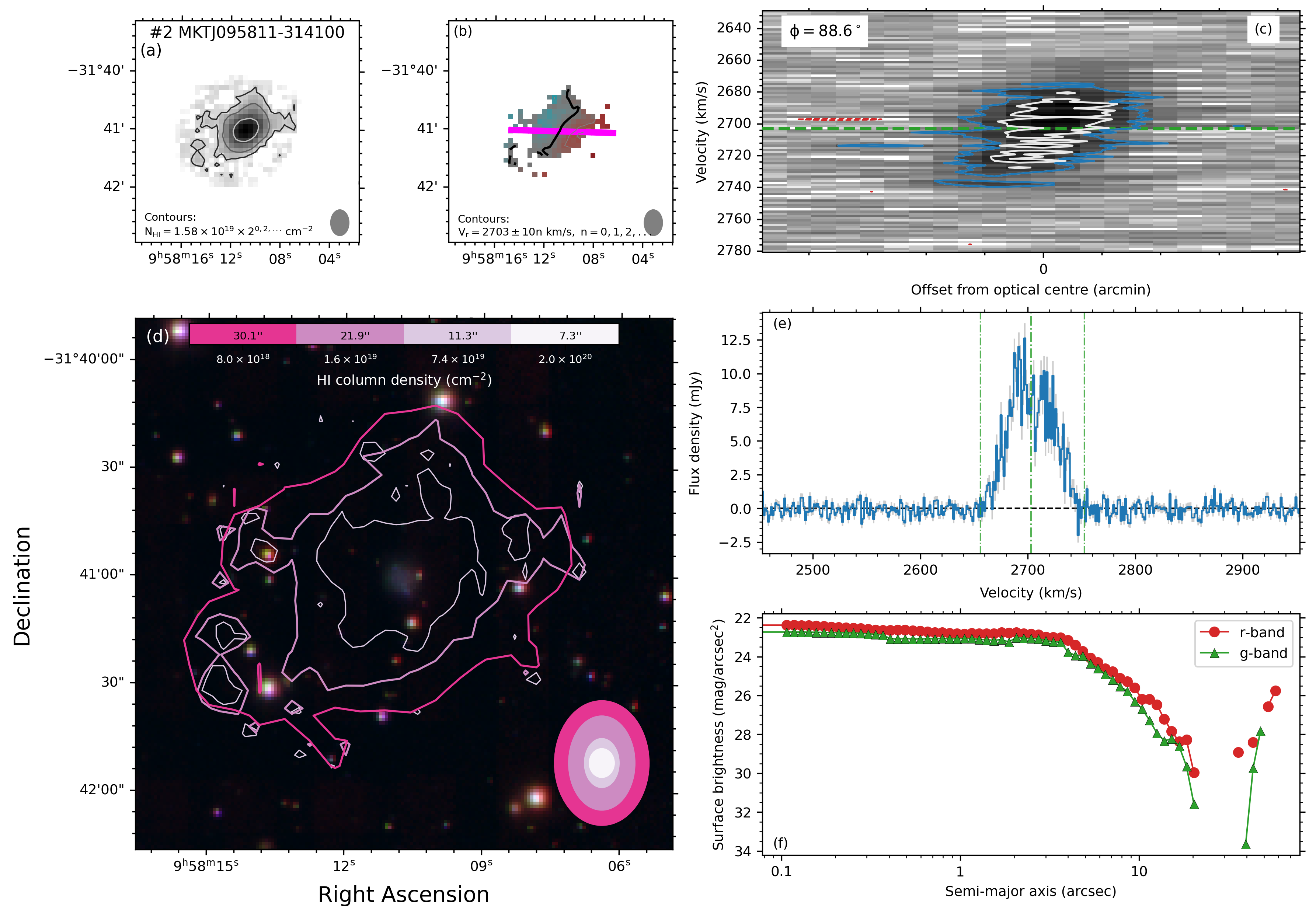}}\\
   \caption{\HI\ and optical data products for each \HI\ detected galaxy. {\em (a)} Total integrated intensity map and {\em (b)} velocity field from the $32\arcsec \times 23\arcsec$ cube and {\em (c)} position velocity slice extracted across the major axis (magenta line). {\em (d)} RGB cutout from Fig.~\ref{fig:fullFieldCont} of the galaxy with N$_{\HI} = 3\sigma$ contour overlaid. {\em (e)} Integrated spectrum of target galaxy. {\em (f)} $r,g$-band surface brightness profiles represented by the red circles and green triangles respectively.}
	\label{fig:atlas}%
\end{figure*} 

\begin{figure*}[p]
	\centering

	\rotatebox{90}{\includegraphics[height=0.45\textwidth]{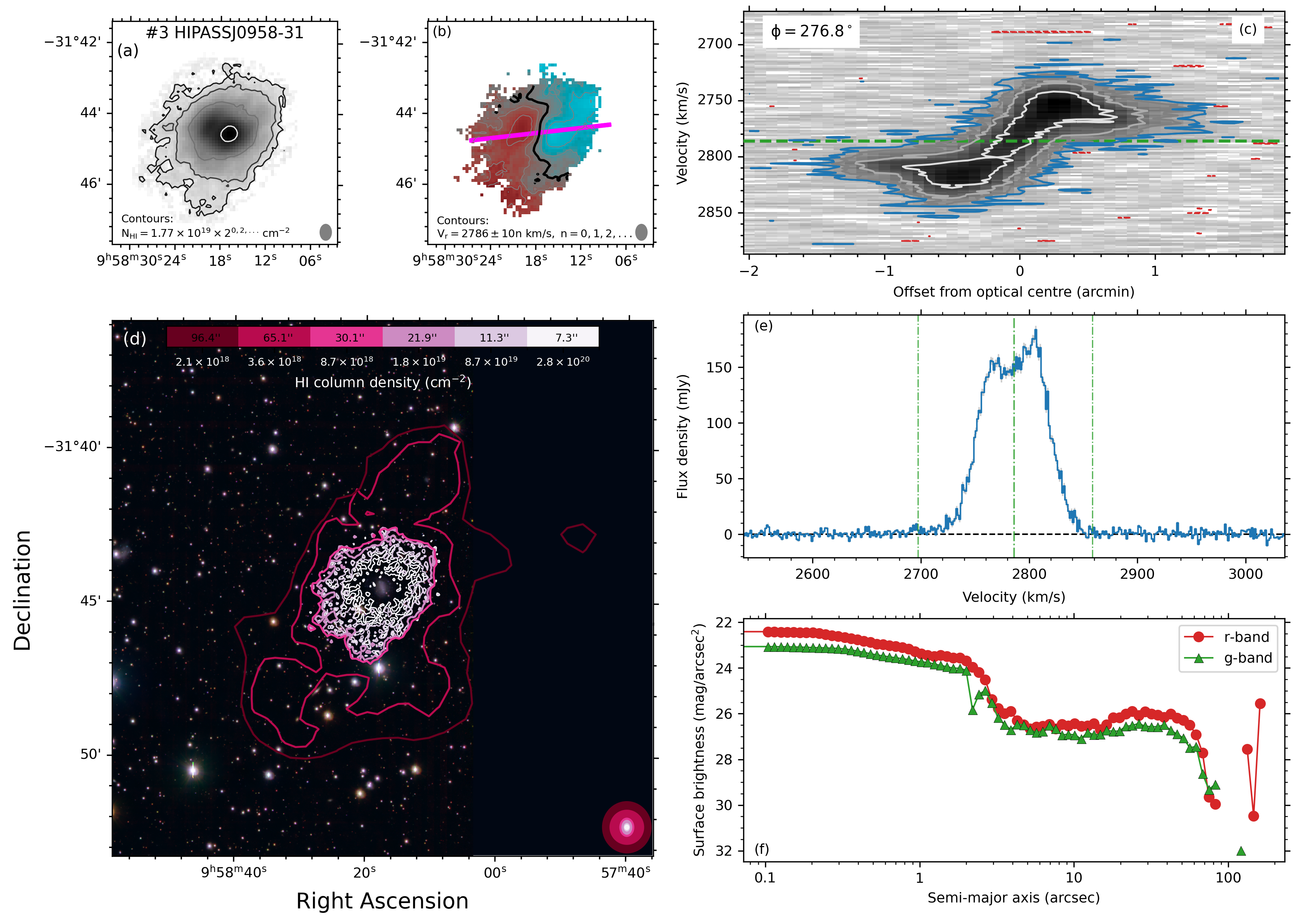}}
	\rotatebox{90}{\includegraphics[height=0.45\textwidth]{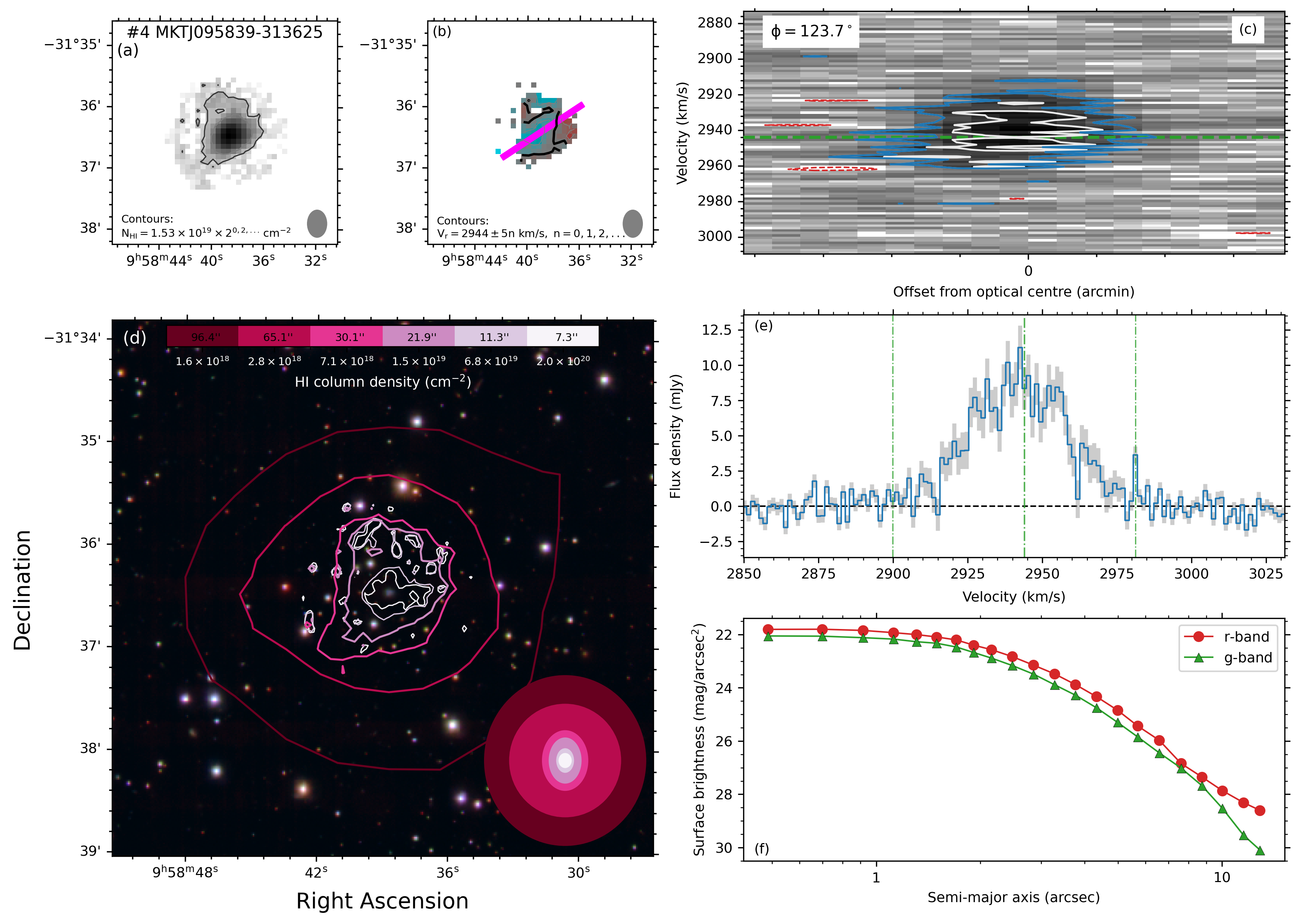}}
	\rotatebox{90}{\includegraphics[height=0.45\textwidth]{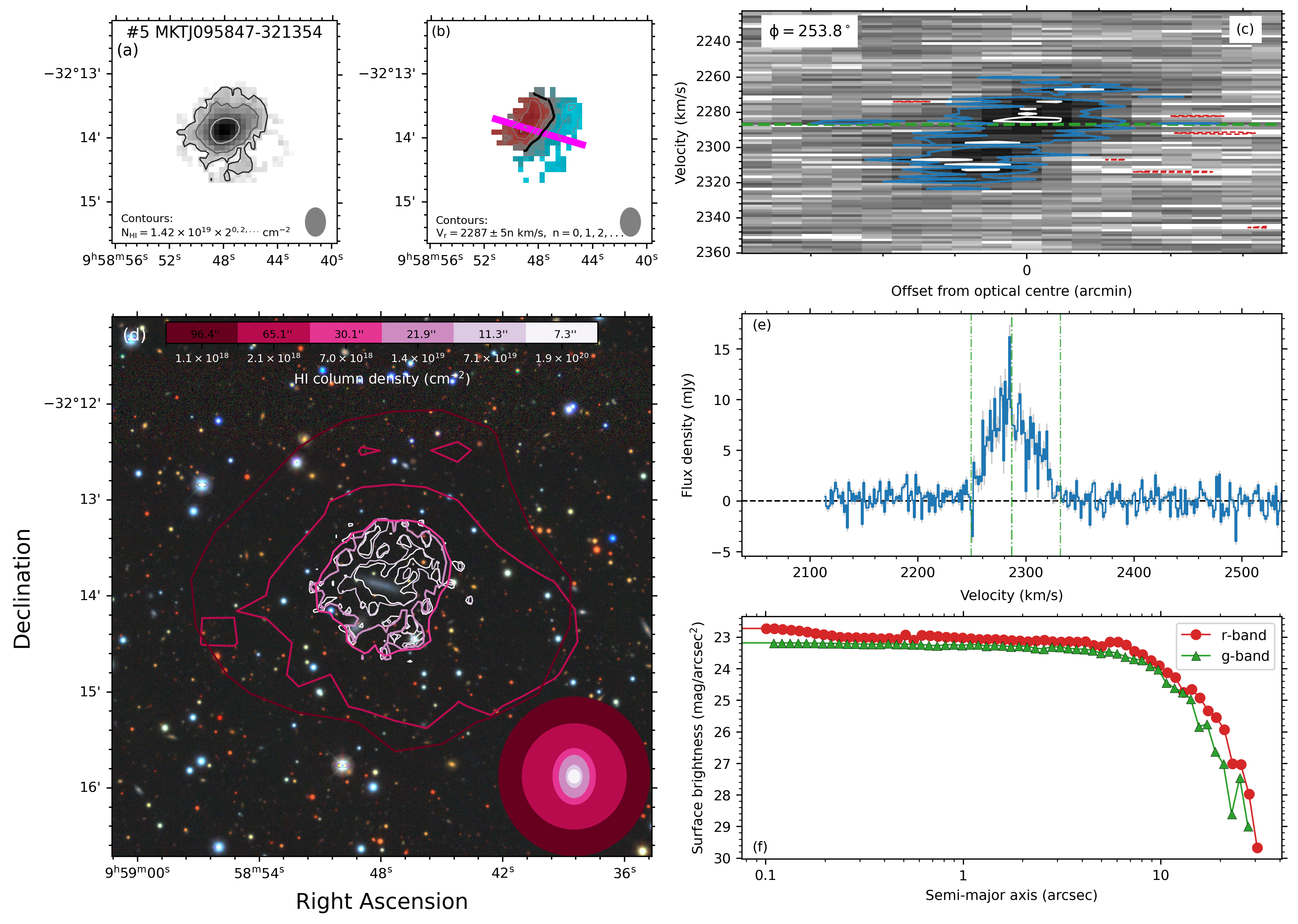}}
    \rotatebox{90}{\includegraphics[height=0.45\textwidth]{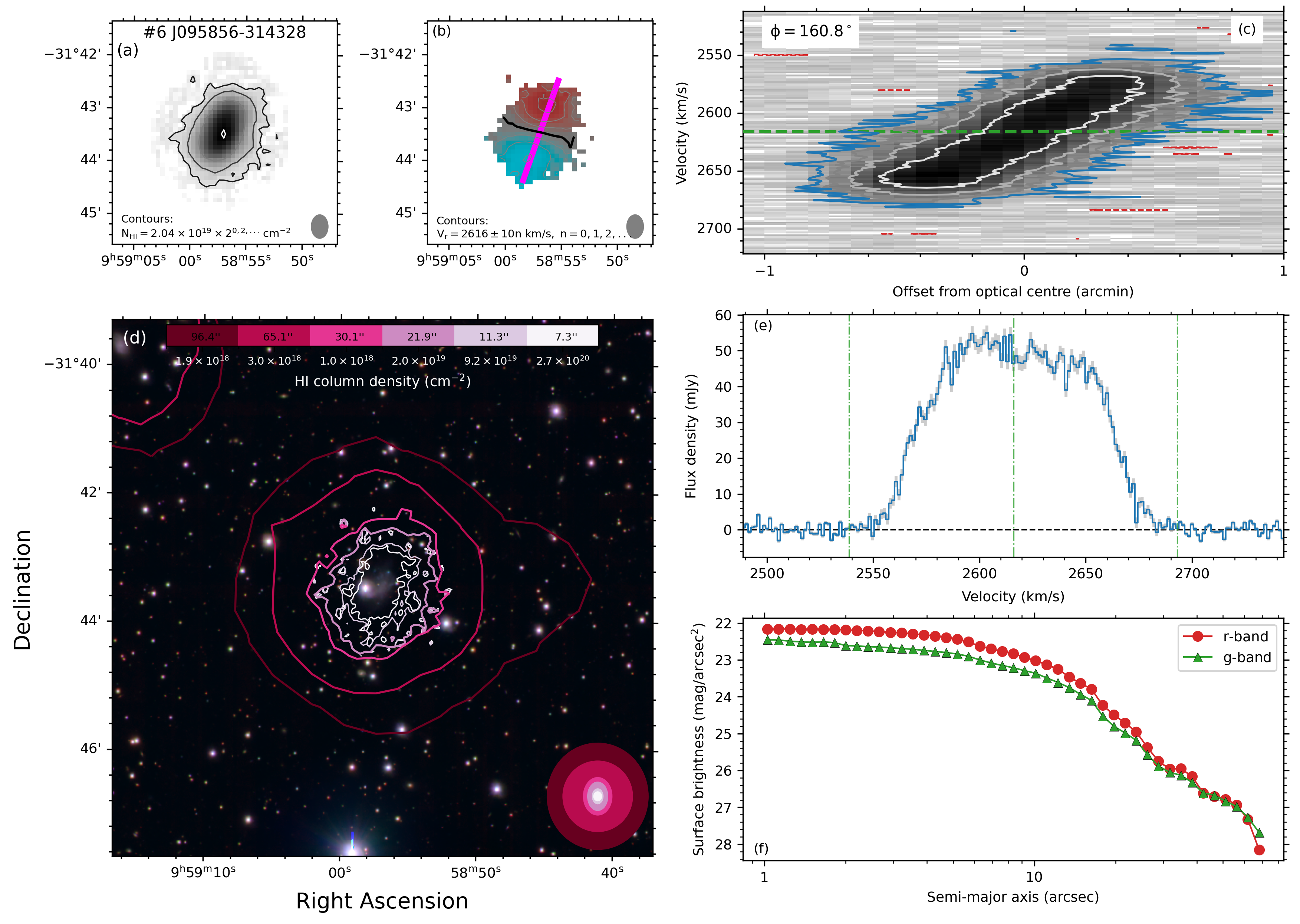}}\\
    \caption{Continued from previous page.}
\end{figure*} 

\begin{figure*}[p]
    \ContinuedFloat
	\centering

	\rotatebox{90}{\includegraphics[height=0.45\textwidth]{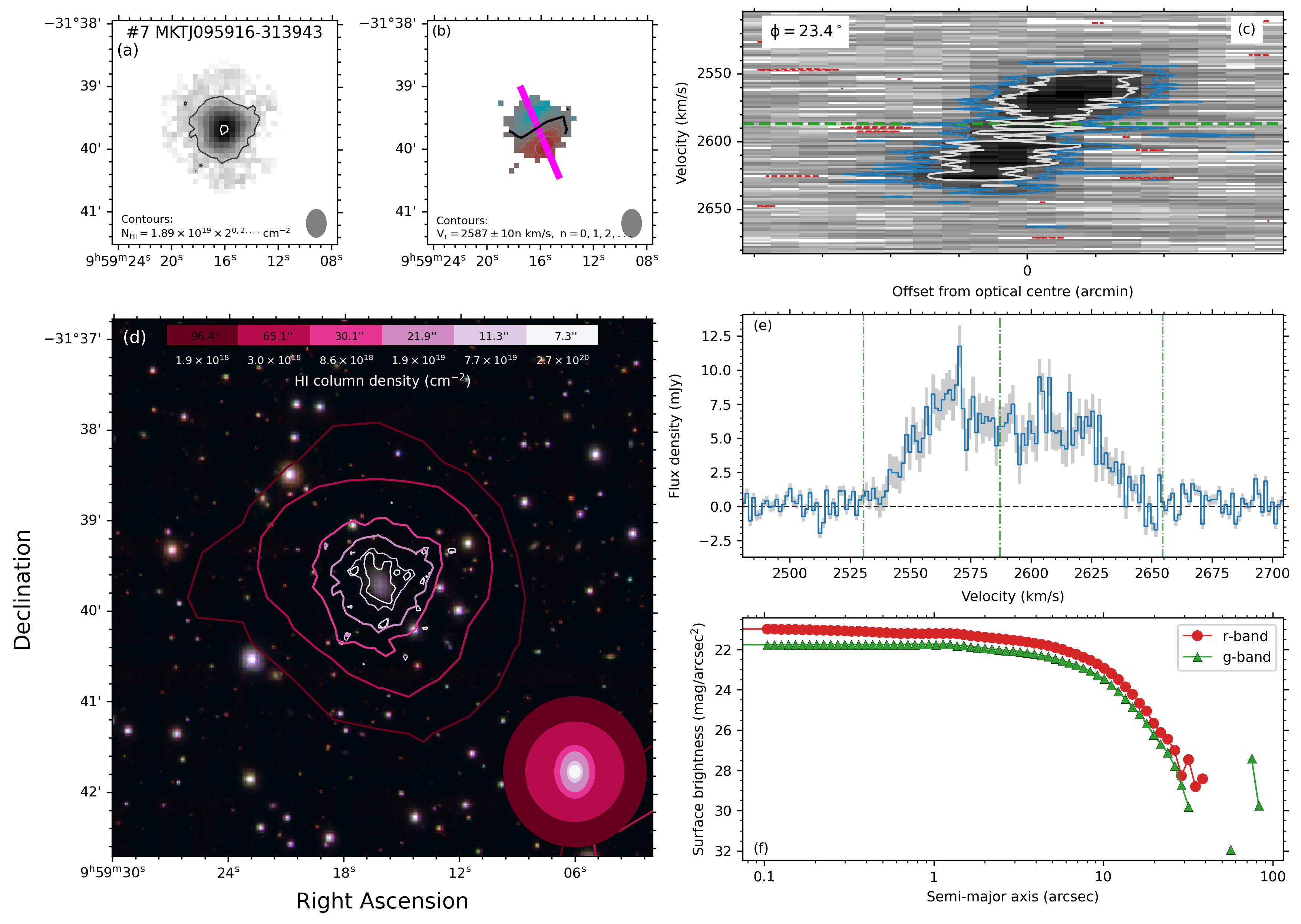}}
	\rotatebox{90}{\includegraphics[height=0.45\textwidth]{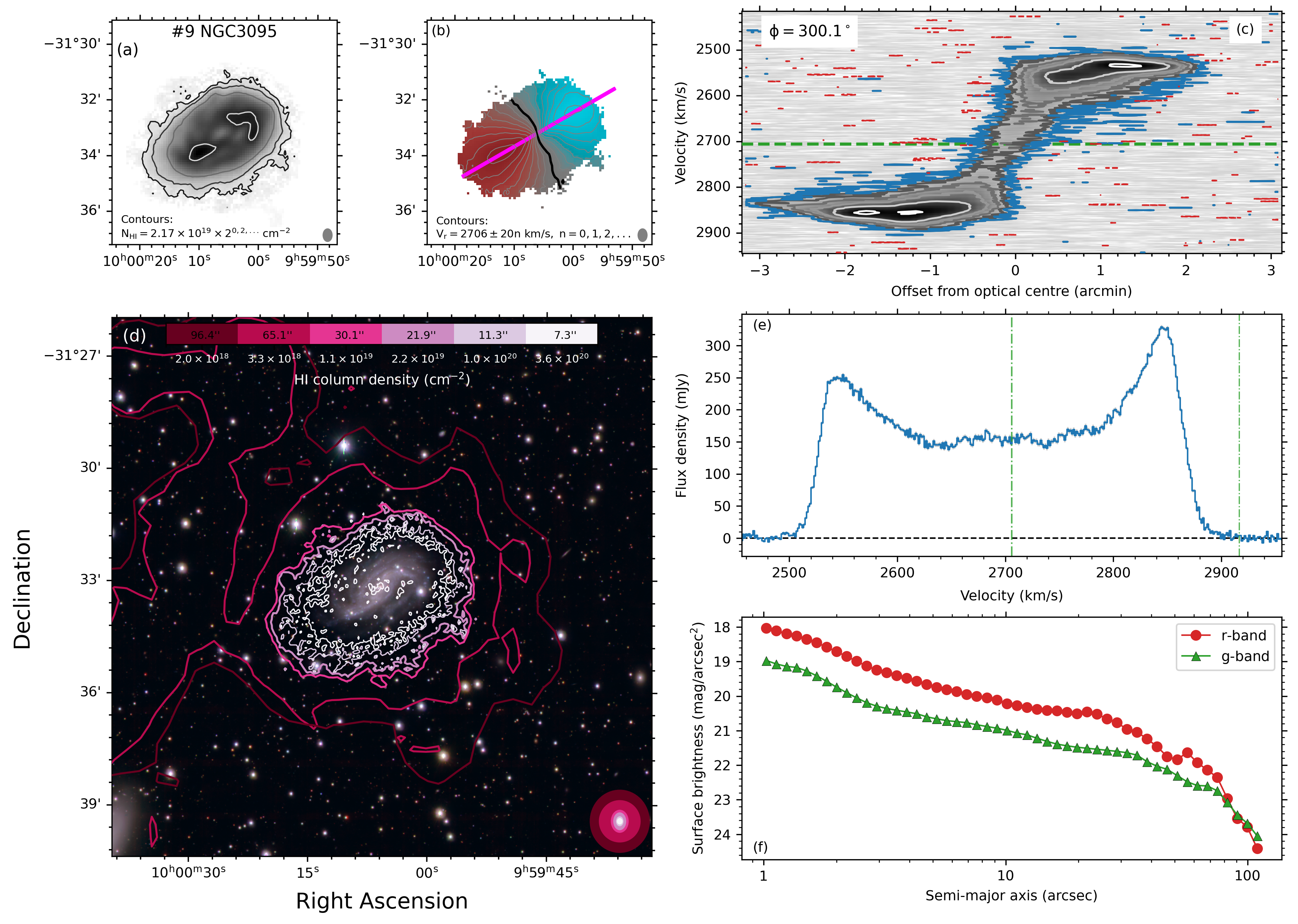}}
    \rotatebox{90}{\includegraphics[height=0.45\textwidth]{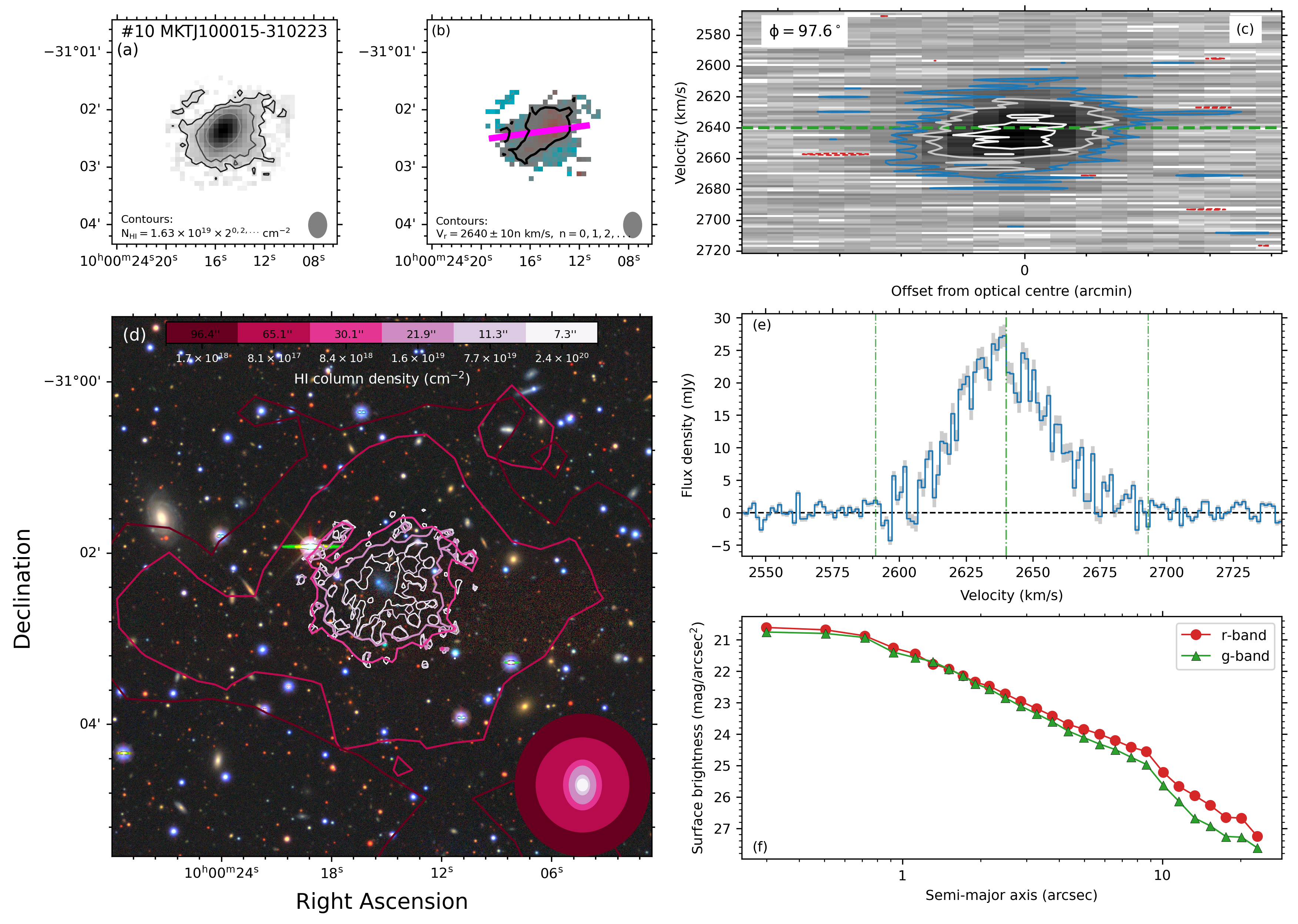}}
    \rotatebox{90}{\includegraphics[height=0.45\textwidth]{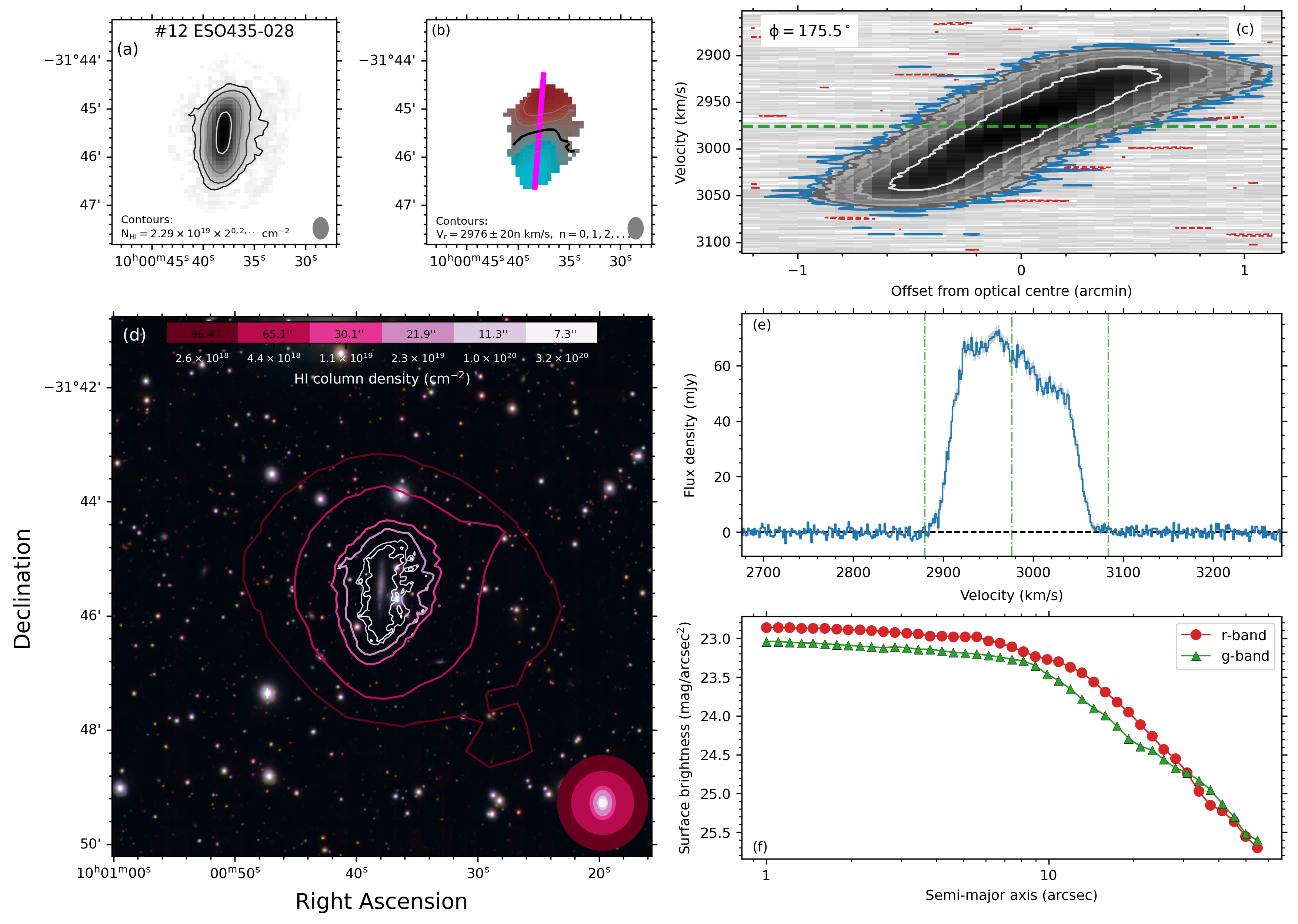}}\\
           \caption{Continued from previous page.} 
\end{figure*} 

\begin{figure*}[p]
    \ContinuedFloat
	\centering
	\rotatebox{90}{\includegraphics[height=0.45\textwidth]{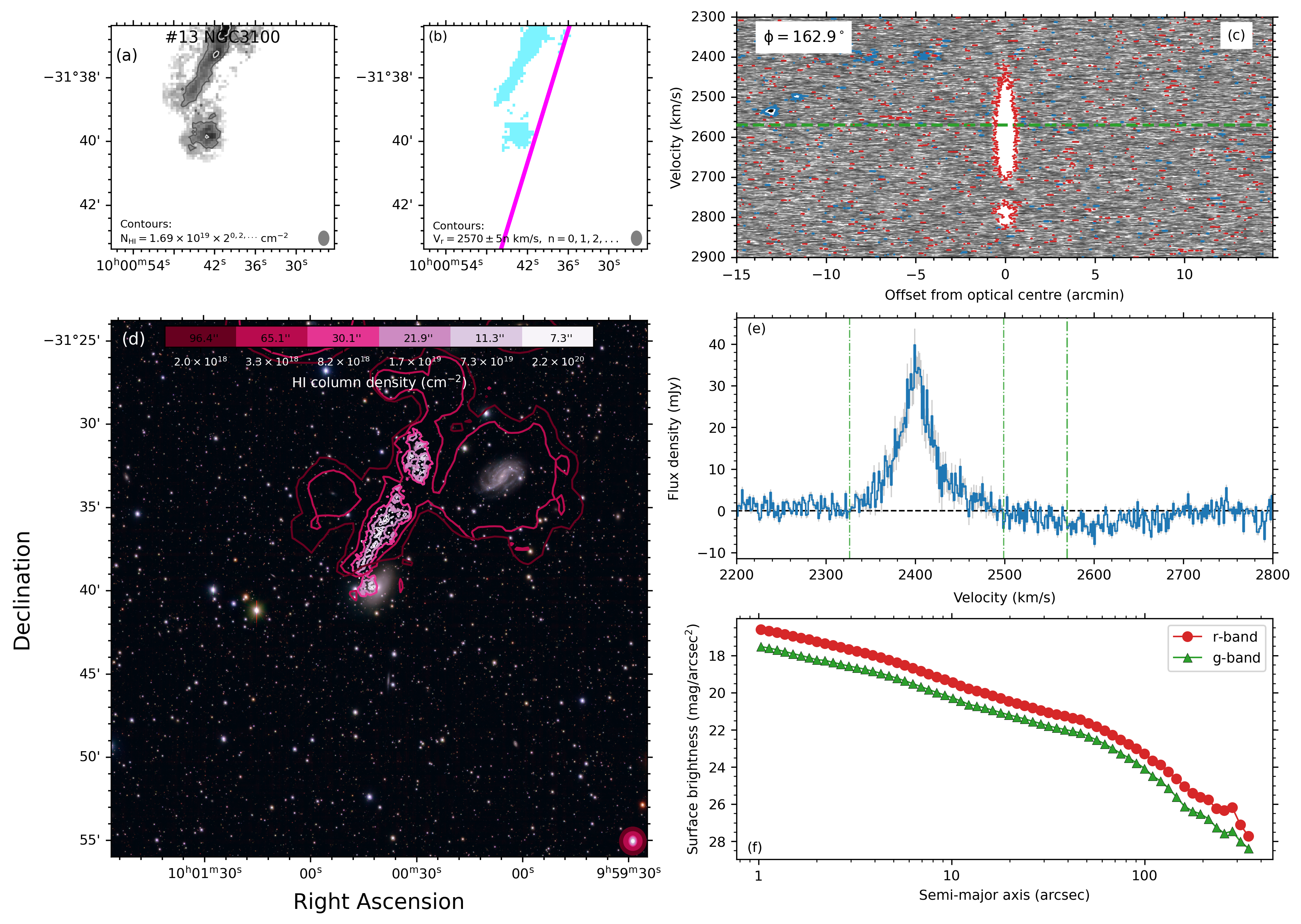}}
	\rotatebox{90}{\includegraphics[height=0.45\textwidth]{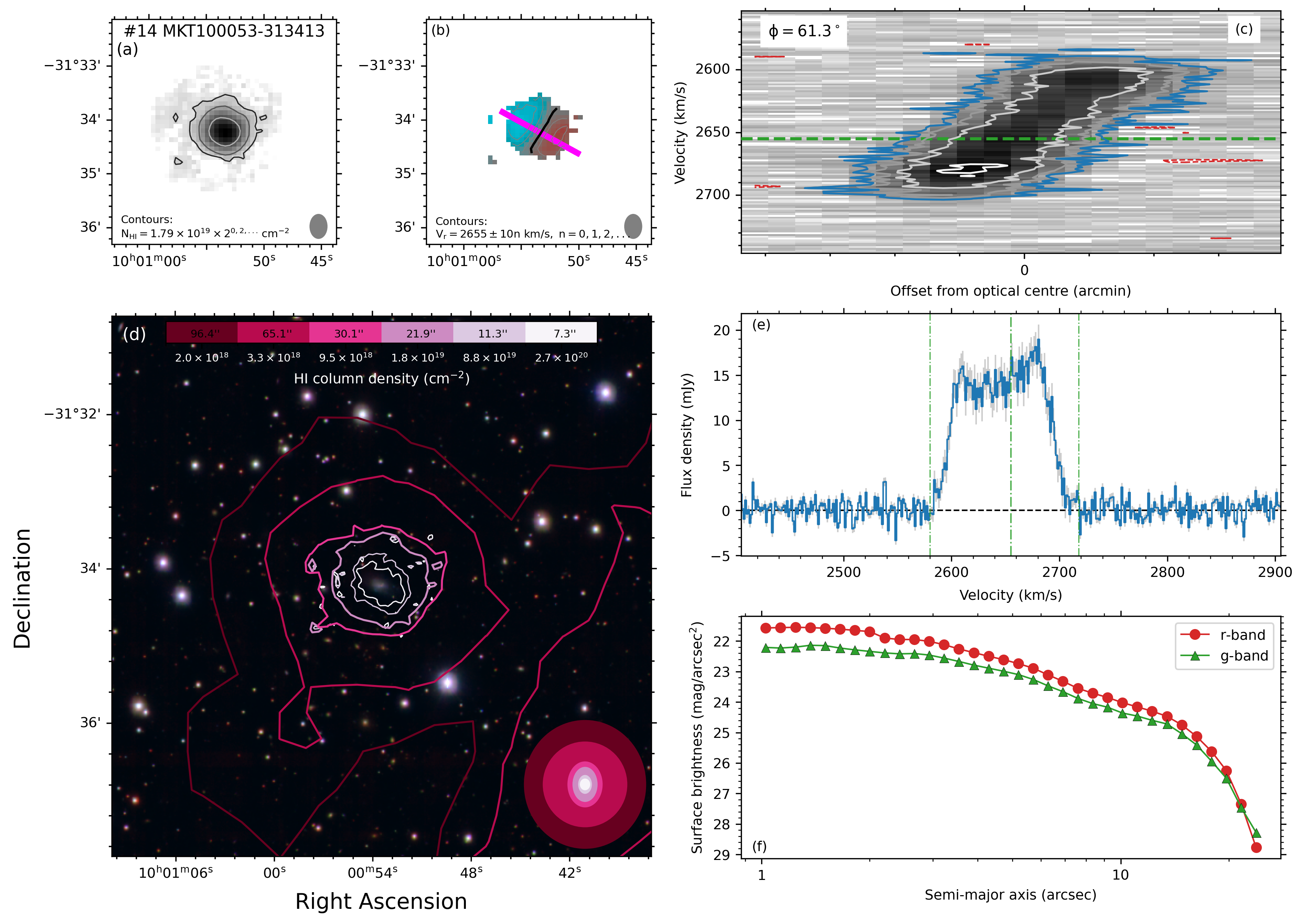}}
        \rotatebox{90}{\includegraphics[height=0.45\textwidth]{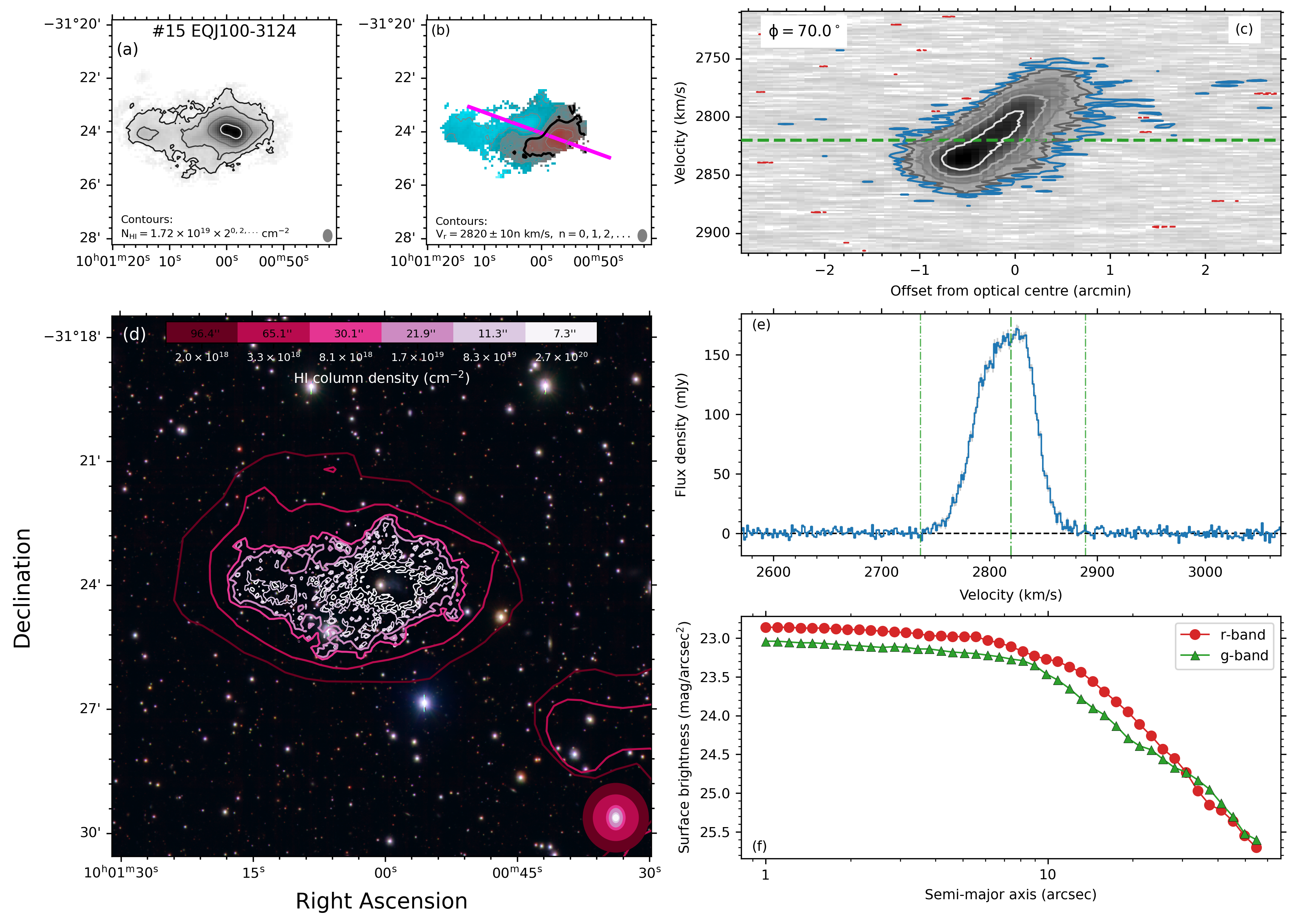}}
    \rotatebox{90}{\includegraphics[height=0.45\textwidth]{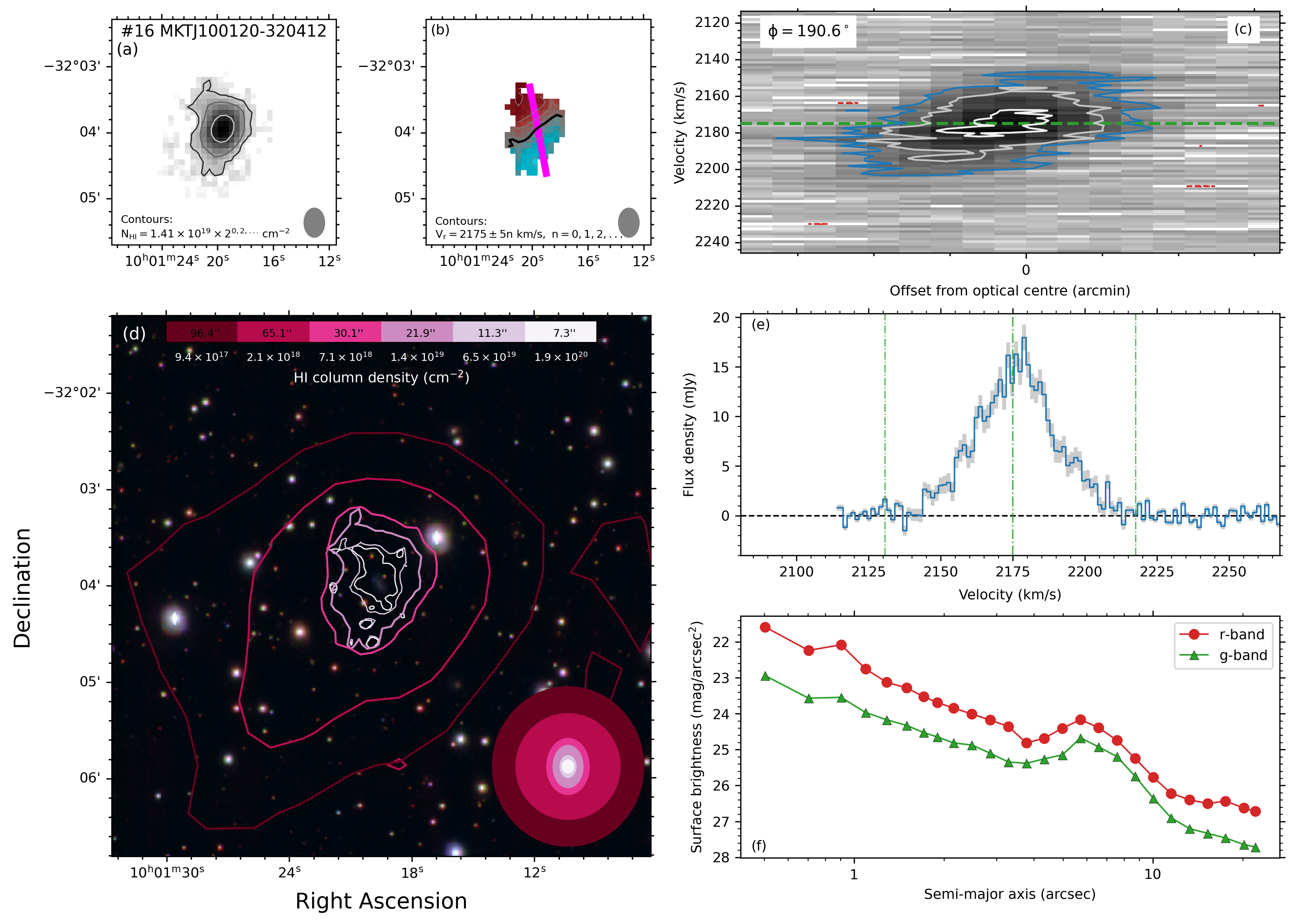}}\\
           \caption{Continued from previous page.} 
\end{figure*} 
    
\begin{figure*}[p]
    \ContinuedFloat
	\centering

	\rotatebox{90}{\includegraphics[height=0.45\textwidth]{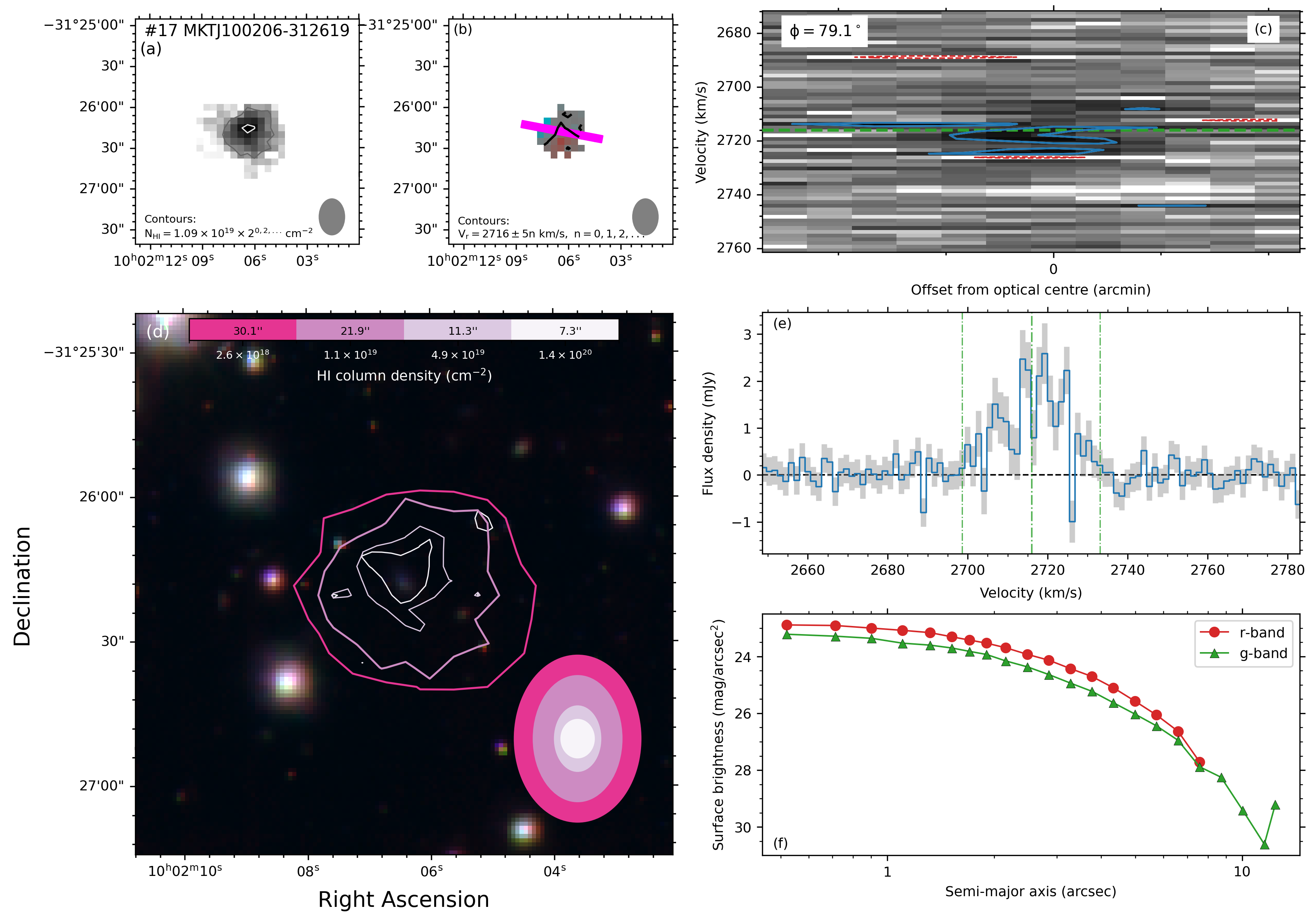}}
	\rotatebox{90}{\includegraphics[height=0.45\textwidth]{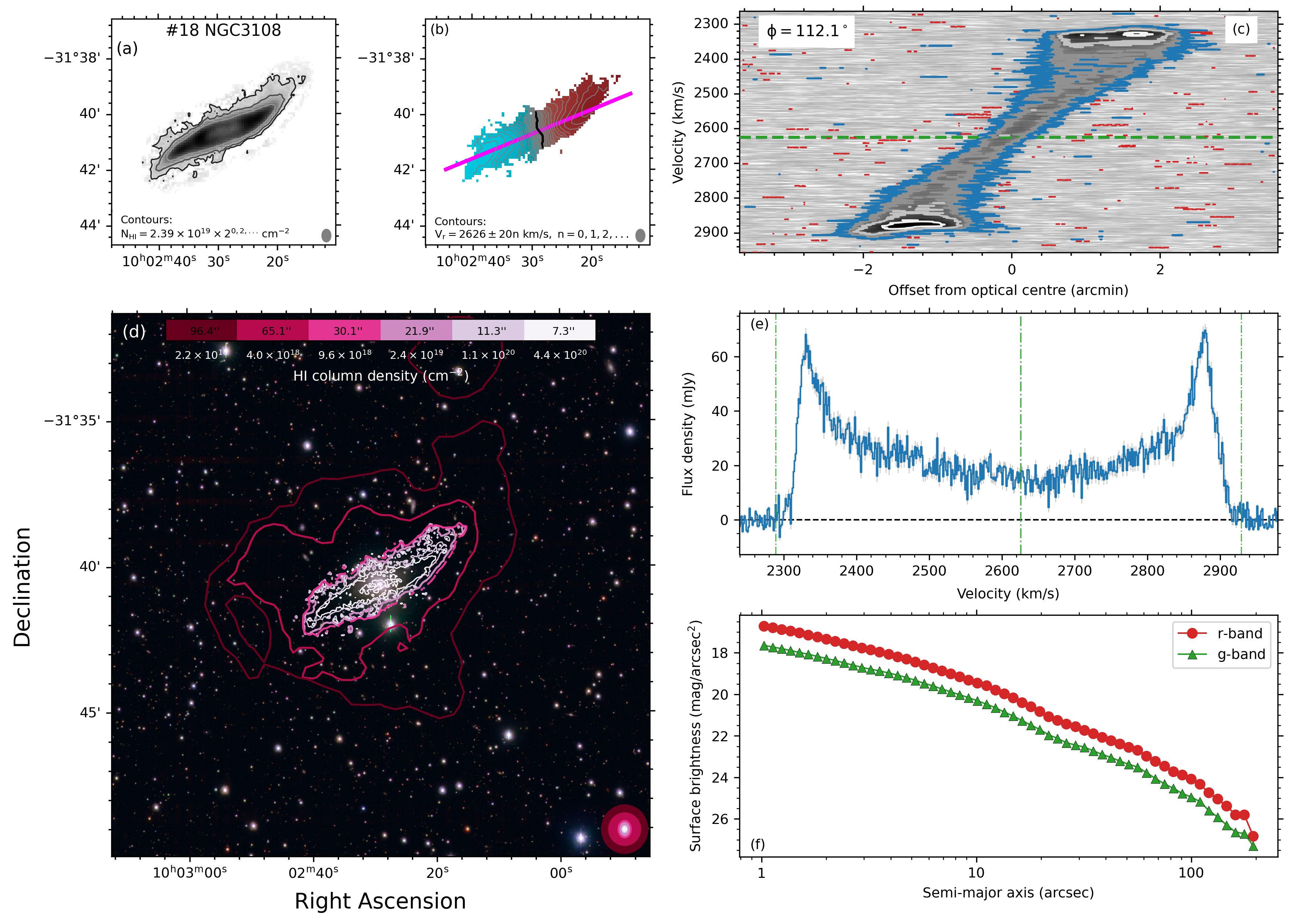}}
        \rotatebox{90}{\includegraphics[height=0.45\textwidth]{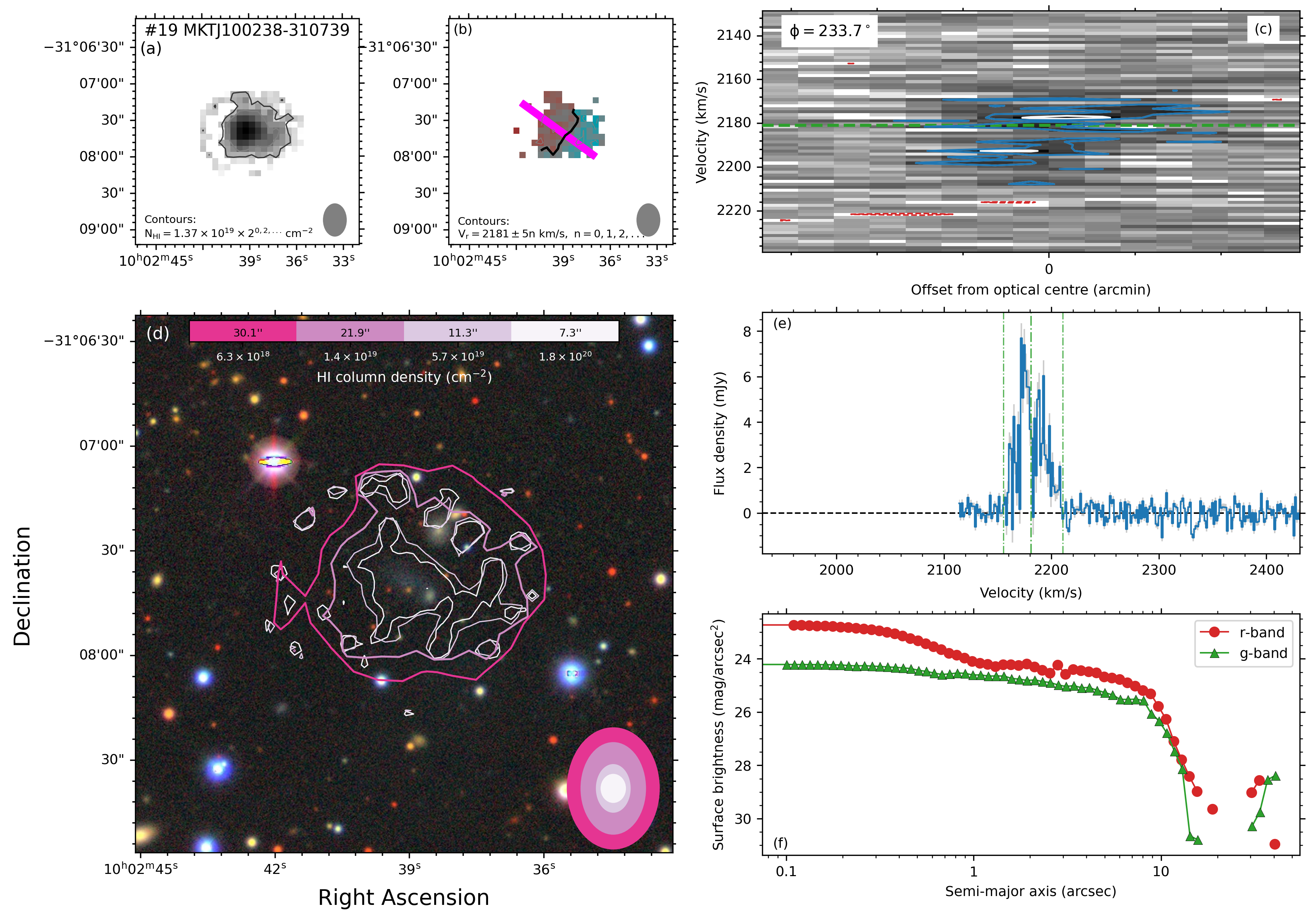}}
    \rotatebox{90}{\includegraphics[height=0.45\textwidth]{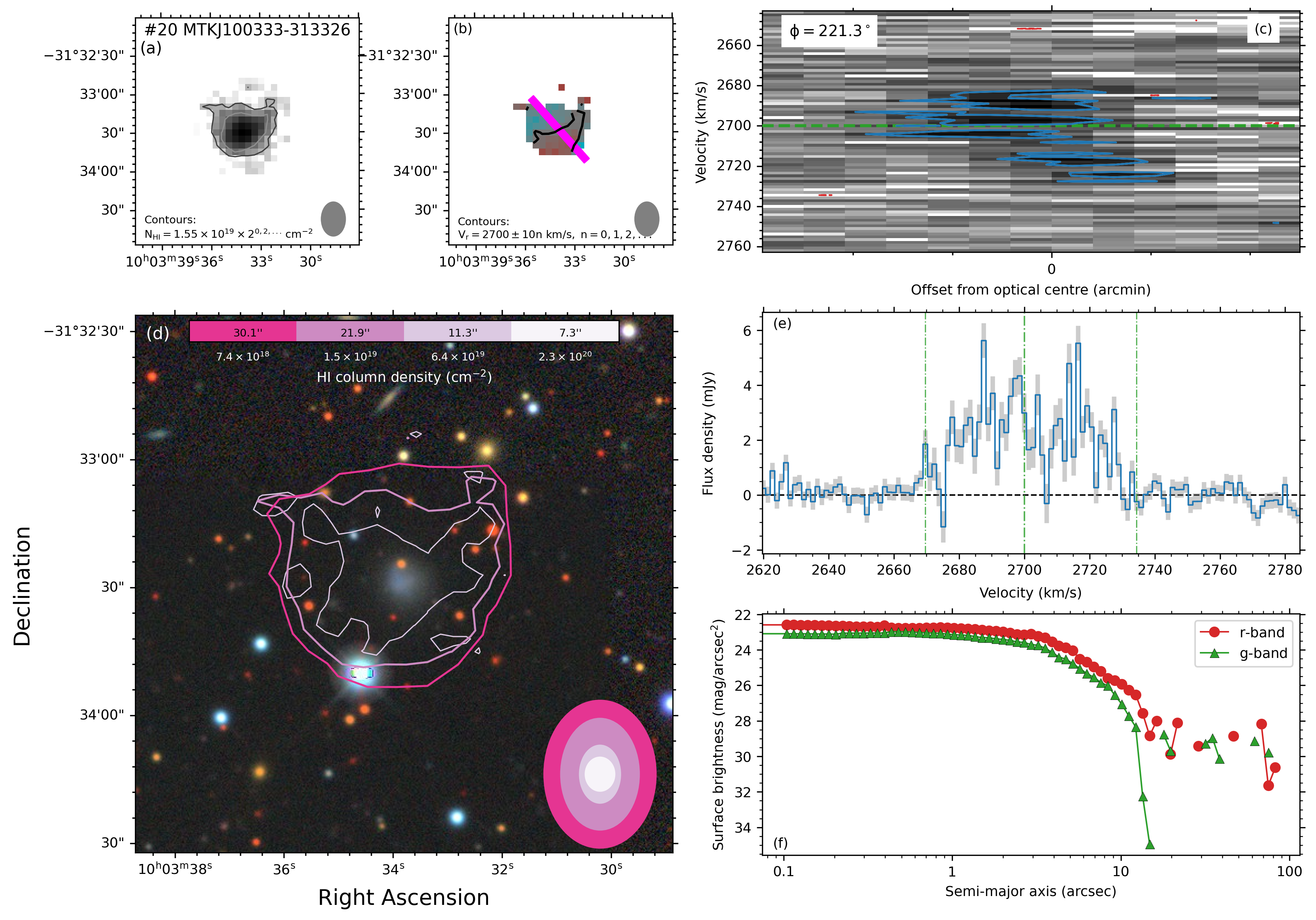}}
           \caption{Continued from previous page.} 
\end{figure*}

\end{document}